\documentclass[preprint]{imsart}
\RequirePackage{amsthm,amsmath,amssymb,bbm}
\RequirePackage[numbers]{natbib}
\RequirePackage[colorlinks,citecolor=blue,urlcolor=blue]{hyperref}

\startlocaldefs
\numberwithin{equation}{section}
\theoremstyle{plain}
\newtheorem{theorem}{Theorem}[section]
\newtheorem{assume}{Assumption}
\newtheorem{lemma}{Lemma}
\newtheorem{cor}{Corollary}[section]
\newtheorem{prop}{Proposition}[section]
\newtheorem{rmk}{Remark}
\DeclareMathOperator*{\argmin}{arg\,min}

\endlocaldefs

\allowdisplaybreaks

\begin{document}
	\begin{frontmatter}
\title{High-dimensional prediction for count response via sparse exponential weights}
		\runtitle{Sparse prediction for count response}
		
\begin{aug}
\author{\fnms{The Tien}~\snm{Mai}\ead[label=e1]{the.t.mai@ntnu.no}\orcid{0000-0002-3514-9636}}
			\address{
	Department of Mathematical Sciences, 
	\\
Norwegian University of Science and Technology,	Trondheim 7034, Norway.
				\\
				\printead[presep={\ }]{e1}
			}
			\runauthor{T.T. Mai}
		\end{aug}
		
		\begin{abstract}
Count data is prevalent in various fields like ecology, medical research, and genomics. In high-dimensional settings, where the number of features exceeds the sample size, feature selection becomes essential. While frequentist methods like Lasso have advanced in handling high-dimensional count data, Bayesian approaches remain under-explored with no theoretical results on prediction performance. This paper introduces a novel probabilistic machine learning framework for high-dimensional count data prediction. We propose a pseudo-Bayesian method that integrates a scaled Student prior to promote sparsity and uses an exponential weight aggregation procedure. A key contribution is a novel risk measure tailored to count data prediction, with theoretical guarantees for prediction risk using PAC-Bayesian bounds. Our results include non-asymptotic oracle inequalities, demonstrating rate-optimal prediction error without prior knowledge of sparsity. We implement this approach efficiently using Langevin Monte Carlo method. Simulations and a real data application highlight the strong performance of our method compared to the Lasso in various settings.
		\end{abstract}
		
		
		
	\end{frontmatter}

\section{Introduction}
\label{sc_intro}

Count data are frequently encountered in various fields such as ecology \cite{richards2008dealing}, medical research \cite{wang2016penalized}, and genomics \cite{lehman2019penalized}, with Poisson regression being a common method for modeling these data. However, in practice, overdispersion—where the observed variance is greater than predicted by the Poisson model—often arises due to factors like missing predictors, correlations among observations, or outliers. To address this, the negative binomial (NB) distribution is often used as an alternative model. Both the Poisson and NB distributions fall within the generalized linear model (GLM) framework \cite{mccullagh1989generalized,tutz2011regression}. In fields like genomics, economics, and finance, high-dimensional data—where the number of features (\(d\)) exceeds the sample size (\(n\))—introduces the ``curse of dimensionality". Analyzing such data requires advanced statistical techniques, including dimensionality reduction through the selection of a sparse subset of significant features. This is essential for effectively analyzing high-dimensional datasets and has driven the development of new methods and theories in statistics \cite{hastie2009elements,buhlmann_vandegeer,wainwright2019high,giraud2021introduction}.

Feature selection techniques for count data have been employed in practice for some time. For instance, \cite{wang2016penalized} and \cite{lehman2019penalized} extended Lasso (Least Absolute Shrinkage and Selection Operator) to Poisson and NB regression, respectively, to analyze prolonged hospital stays after pediatric cardiac surgery and micronuclei frequencies in genomics studies. More recently, \cite{li2022heterogeneous} proposed a heterogeneous NB model, using double NB regressions to account for both the dependent variable and the overdispersion parameter, applying double Lasso techniques for model selection in both regressions.
Numerous studies have examined estimation errors, including the works of \cite{blazere2014oracle,zhang2022elastic,peng2024oracle}, while \cite{abramovich2016model,abramovich2018high} offer theoretical insights into prediction risk.

On the other hand, Bayesian approaches for modeling count responses in high-dimensional settings have received relatively limited attention. The primary challenge lies in the lack of conjugacy when using Poisson or negative binomial regression models, which complicates the computation. Recently, an adaptive Metropolis-Hastings sampling method has been introduced for Bayesian Poisson regression with sparsity in \cite{dAngelo2023efficient}. Additionally, a more recent development is the proposal of a Variational Bayesian approach for sparse Poisson regression \cite{kharabati2023variat}. Some studies have also explored the frequentist properties of the posterior in Bayesian methods for generalized linear models, as seen in \cite{jiang2007bayesian,jeong2021posterior}. However, to the best of our knowledge, there is no theoretical results on prediction for Bayesian approaches with count response data.

To address the existing gap in the field, our paper introduces a novel methodological framework aimed at overcoming these challenges. We employ a probabilistic machine learning approach, incorporating an exponential weight aggregation procedure and presenting a pseudo-Bayesian method specifically designed for high-dimensional count data prediction. Our pseudo-Bayesian approach extends the traditional Bayesian framework by evaluating data fit through a broader notion of risk or quasi-likelihood, rather than strictly relying on the likelihood. This flexible approach, which has gained traction in generalized Bayesian inference and machine learning \cite{bissiri2013general,grunwald2017inconsistency,kno2019}, allows us to avoid restrictive assumptions about the data-generating process and enables a sharper focus on specific objectives such as prediction rather than estimation. We further enhance our method by employing a scaled Student prior distribution \cite{dalalyan2012mirror,dalalyan2012sparse} to promote sparsity. A key contribution of our work is the development of a novel loss/risk measure tailored specifically for count data prediction.

By utilizing PAC-Bayesian bounds technique \citep{STW, McA, catonibook}, we establish prediction risk bounds for our method under varying conditions. Central to our findings are non-asymptotic oracle inequalities, which demonstrate that the prediction error of our approach closely matches the optimal error, providing solid theoretical assurances of its efficacy. Specifically, we derive results under different loss function conditions, showing that even with basic assumptions like bounded loss, a slow rate can be achieved. Under more stringent assumptions, a fast rate is also attainable. We further explore the case of unbounded loss. In all scenarios, our results exhibit adaptability to the unknown sparsity level of the true predictor. While the PAC-Bayesian framework has been successfully employed in various sparse settings \citep{dalalyan2008aggregation, alquier2011PAC, dalalyan2012sparse, alquier2013sparse, guedj2013pac, ridgway2014pac,mai2015,guedj2018pac, luu2019pac, mai2023high}, its application to predicting count data has not been studied before. For an in-depth review of PAC-Bayes bounds and recent advancements, refer to \cite{guedj2019primer, alquier2024user}.

The concept that scaled Student priors yield optimal rates in high-dimensional regression was originally introduced by \cite{dalalyan2008aggregation,dalalyan2012mirror,dalalyan2012sparse}. These priors have since been applied in other contexts such as binary classification \cite{mai2023high} and quantile regression \cite{mai2024sparse}. Although not conjugate for our specific problem, the scaled Student prior proves advantageous for implementing a gradient-based sampling technique. We introduce a Langevin Monte Carlo (LMC) algorithm to sample from the pseudo-posterior and compute the mean estimator, offering an efficient sampling approach. This method ensures effective exploration of the parameter space while remaining computationally manageable. Comparative studies with the Lasso method, in both simulations and real data application, reveal that our approach delivers strong performance across various models, such as Poisson regression and negative binomial regression, underscoring its robustness and flexibility in addressing high-dimensional and overdispersed count data.

The remainder of the paper is organized as follows. In Section \ref{sc_problem_method}, we introduce the problem and present our proposed methodology. Section \ref{sc_theory} outlines the main theoretical results. Section \ref{sc_numberical} details the implementation of our method, including simulation studies and an application to real data. The technical proofs are provided in Appendix \ref{sc_proofs}, and the paper concludes with final remarks in Section \ref{sc_conlsution}.

\section{Problem and method}
\label{sc_problem_method}
\subsection{Problem statement}
Consider a collection of \( n \) independent and identically distributed (i.i.d.) random variables \( (Y_i, X_i) \in \mathbb{N} \times \mathbb{R}^d \), for \( i = 1, \dots, n \), sampled from the distribution \( \mathbf{P} \). Suppose $ \theta \in \mathbb{R}^d $ is a linear predictor and consider the prediction problem: that given $ (Y, X) \sim \mathbf{P} $, the prediction for $ Y $ is made by $ e^{ X^\top \theta} $. The ability of this predictor to predict a new $ Y $ is then assessed by the following risk
\begin{align}
\label{eq_new_count_risk}
R (\theta) = \mathbb{E}_{\mathbf{P}}
[ (Y - \exp (X^\top \theta) )^2 ]
\end{align}
 and the empirical  risk is given as
\begin{align*}
r_n (\theta) 
= 
\frac{1}{n} \sum_{i=1}^{n}
(Y_i - \exp (X_i^\top \theta) )^2
.
\end{align*}

Define 
$$
\theta^* 
= \argmin_{ \theta\in\mathbb{R}^d } R (\theta), 
\,\, 
\text{and} 
\quad
R^* := \min_{ \theta\in\mathbb{R}^d } R (\theta)
.
$$
We consider in this paper the prediction problem, i.e., build an estimator $ \hat{\theta} $ from the data $ (Y_i,X_i)_{i=1,\ldots, n } $ such that $  R(\hat{\theta}) $ is close to $ R^* $ up to a positive remainder term as small as possible.

A high-dimensional sparse scenario is studied in this paper: we assume that $ s^* < n < d $, where $ s^*:= \|\theta^* \|_0 $.

The idea of expressing the predictor in exponential form and defining the risk as in \eqref{eq_new_count_risk} is inspired by certain regression models for count response data. For instance, in Poisson regression with a log-link function for the covariate \( X \), as discussed in \cite{wang2016penalized,lehman2019penalized}, the conditional expectation of \( Y \) given \( X \) is \( \mathbb{E}(Y | X) = \exp(X^\top \theta) \). This relationship holds similarly for Negative Binomial regression, as shown in \cite{li2022heterogeneous,zhang2022elastic,zilberman2024high}. In fact, through simulations, we demonstrate that this loss performs effectively for both Poisson regression and Negative Binomial regression models. Initially, we had experimented with using the log-likelihood as the loss function in both the Poisson and Negative Binomial regression models. However, the results from those simulations were not as promising as those obtained with the current loss function. The performance improvements we have achieved with the current approach suggest that our chosen loss function is better suited for these models, yielding more accurate and reliable outcomes across various scenarios.

\subsection{A sparse EWA approach}

For any \(\lambda > 0\), with a prior distribution $ \pi $ given below, we consider the following Gibbs posterior \(\hat{\rho}_{\lambda} \),
\begin{equation}
\label{eq_Gibbs_poste}
\hat{\rho}_{\lambda} (\theta) 
\propto 
\exp[-\lambda r_n(\theta)] \pi(\theta)
, 	
\end{equation}
and let $ \hat{\theta} = \int \theta \hat{\rho}_{\lambda} ({\rm d}\theta)  $ be our mean estimator.

The Gibbs posterior in \eqref{eq_Gibbs_poste} is also known as the EWA (exponentially weighted aggregate) procedure, as referenced in \cite{alquier2016,catonibook,dalalyan2012sparse,dalalyan2008aggregation}. The use of $\hat{\rho}_\lambda $ is driven by the minimization problem outlined in Lemma \ref{lemma:dv}, rather than strictly following traditional Bayesian methods. Importantly, there is no requirement for a likelihood function or a complete model; the focus is solely on the empirical risk derived from the loss function. In this paper, we consistently use $\pi$ to represent the prior and $\hat{\rho}_\lambda$ to denote the pseudo-posterior. The purpose of the EWA is to adjust the distribution to favor parameter values with lower in-sample  empirical risk, with the degree of adjustment governed by the tuning parameter $\lambda$, which will be explored further in the following sections.

The choice of prior distribution plays a critical role in ensuring an accurate prediction error rate in high-dimensional models. 
For a fixed constant \( C_1 >0 \), for all $ \theta \in \mathbb{R}^d  $ that \(  \|\theta\|_1 \leq C_1 \), we adopt the scaled Student distribution as our prior distribution.
\begin{eqnarray}
\label{eq_priordsitrbution}
\pi (\theta) 
\propto 
\prod_{i=1}^{d} 
(\varsigma^2 + \theta_{i}^2)^{-2}
,
\end{eqnarray}
where $ \varsigma>0 $ is a tuning parameter.  This prior has been applied in various sparse settings, as demonstrated in \cite{dalalyan2012mirror, dalalyan2012sparse, mai2023high}. In this context, $C_1$ acts as a regularization parameter and is generally assumed to be very large. As a result, the distribution of $\pi $ approximates that of $S\varsigma \sqrt{2}$, where $S$ is a random vector with i.i.d. components drawn from a Student's t-distribution with 3 degrees of freedom. By choosing a very small value for $\varsigma$, the majority of elements in $\varsigma S$ are concentrated near zero. However, because of the heavy-tailed nature, a small proportion of components in $\varsigma S$ significantly deviate from zero. This property allows the prior to effectively encourage sparsity in the parameter vector. The significance of heavy-tailed priors in fostering sparsity has been explored in prior studies as in \citep{seeger2008bayesian, johnstone2004needles, rivoirard2006nonlinear, abramovich2007optimality, carvalho2010horseshoe, castillo2012needles, castillo2015bayesian, castillo2018empirical, ray2022variational}.

\section{Theoretical results}
\label{sc_theory}
\subsection{Assumptions}

Put $ \ell_i (\theta) = (Y_i - \exp (X_i^\top \theta) )^2 $.

We outline the assumptions that are essential for obtaining our theoretical findings.

\begin{assume}[bounded loss]
	\label{assume_bounded_loss}
	We assume that there exists a constant $ 0 < C < \infty  $ such that
	$ | \ell_i (\theta) - \ell_i (\theta^*) |
	 \leq C $.
\end{assume}

\begin{assume}[random design]
	\label{assume_X_bounded}
	We assume that there exists a constant $ C_{\rm x} >0 $ such that
	$ 	\mathbb{E} \|X_1\| \leq C_{\rm x} < \infty $.
\end{assume}

\begin{assume}[Lipschitz]
	\label{assume_Lipschitz}
	We assume that there exists a constant $ 0 < C_L < \infty  $ such that
	$ \mathbb{E} | \ell_i (\theta) - \ell_i(\theta^*) | \leq C_L \mathbb{E} | X_i^\top (\theta-\theta^*) |  $.
\end{assume}

\begin{assume}[Bernstein's condition]
	\label{assum_bernstein}
	Assume that for any $\theta\in\Theta $, $ R(\theta) \geq R(\theta^*) $  and there is a constant $K>0 $ such that, 
	\begin{equation*}
		\mathbb{E} \{ 
		|  X^\top (\theta - \theta^*) |^2 \}
		\leq 
		K [ R (\theta) - R(\theta^*) ]
		,
	\end{equation*}
\end{assume}

\begin{assume}[Heavy-tailed loss]
	\label{assum_heavy_tailed}
	Assume that for any $\theta\in\Theta $ there are two positive constants $ H_1 > 0, H_2 > 0 $ such that, 
	\begin{align*}
	\mathbb{E} \{ 
	[ \ell_i (\theta)- \ell_i (\theta^*) ]^2 \}
	\leq 
	H_1 [ R (\theta) - R(\theta^*) ]
	,
	\end{align*}
	and for all integers $ k\geq 3 $,
	\begin{align*}
\mathbb{E} \{ 
| \ell_i (\theta)- \ell_i (\theta^*) |^k \}
\leq 
H_1 [ R (\theta) - R(\theta^*) ] \frac{k! H_2^{k-2}}{2}
.
\end{align*}	
\end{assume}

Assumption \ref{assume_bounded_loss} is relatively mild, as it is reasonable to assume that \( Y_i \) and \( X_i \) are almost surely bounded. Under this condition, a slow prediction error rate on the order of \( n^{-1/2} \) can be achieved, as in Theore \ref{thm_main_2}. Assumption \ref{assume_X_bounded} is a commonly adopted condition, frequently employed in both Bayesian approaches \citep{jiang2007bayesian} and non-Bayesian studies \citep{abramovich2018high}. Assumption \ref{assume_Lipschitz} plays a crucial role in all of our results and is analogous to the Lipschitz condition for the loss function. Assumption \ref{assum_bernstein}, can be regarded as a type of Bernstein's condition. This condition has been extensively studied in the learning theory literature, for example in \citep{mendelson2008obtaining,zhang2004statistical,alquier2019estimation,elsener2018robust}. For our purposes, this assumption is not only important to achieving a fast rate but also allows us to obtain some results regarding estimation error regarding the predictor.

Assumption \ref{assum_heavy_tailed} asserts that the relative loss \( \ell_i (\theta) - \ell_i (\theta^*) \) is a sub-gamma random variable. This is a standard condition in the analysis of machine learning algorithms, indicating that the tails of \( \ell_i (\theta) - \ell_i (\theta^*) \) are not heavier than those of a gamma random variable. This class includes, for instance, Gaussian random variables and any random variable that is centered and bounded in absolute value by a constant \( C > 0 \), as discussed in Chapter 2 of \cite{boucheron2013concentration}. Similar assumptions to \ref{assum_heavy_tailed} have been previously employed in \citep{alquier2011PAC,alquier2013sparse,mai2015} to derive PAC-Bayesian bounds for unbounded loss in various regression models with continuous responses.

\subsection{Slow rates}

We first provide non-asymptotic bounds on the excess  risk with minimal assumption.

Put $ \tilde{\xi} = s^* \log ( n\sqrt{d} / s^*)
/\sqrt{n}  $.

\begin{theorem}
	\label{thm_main_2}
Assume that Assumption \ref{assume_bounded_loss}, \ref{assume_X_bounded} and  \ref{assume_Lipschitz} are satisfied.  Take $\lambda= \sqrt{n} $, $ \varsigma = ( C_L C_{\rm x} n\sqrt{d})^{-1} $.  Then for all $ \theta^* $ such that $  \| \theta^*\|_1 \leq C_1 - 2d\varsigma $ we have that
	\begin{equation*} \mathbb{E}\,\mathbb{E}_{\theta\sim\hat{\rho}_{\lambda}} [R(\theta) ]- R^*
		\leq 
		\mathcal{C}_1
		\,	\tilde{\xi} 
		,
	\end{equation*}	
	and	 with probability at least $ 1-\varepsilon, \varepsilon\in (0,1) $ that
	\begin{equation*} \mathbb{E}_{\theta\sim\hat{\rho}_{\lambda}} [R(\theta) ]- R^*
		\leq 
		\mathcal{C}'_1
		[ \tilde{\xi} +	n^{-1/2} \log( 2 / \varepsilon) ]
		,
	\end{equation*}
	for some constant $ \mathcal{C}_1 , \mathcal{C}'_1 > 0 $ depending only on $ C,C_1,C_{\rm x}, C_L $.
\end{theorem}

The proofs are given in Appendix \ref{sc_proofs}, where we rely on the `PAC-Bayesian bounds' approach from \cite{catonibook} as our key technical framework. For a deeper exploration of PAC-Bayes bounds and their recent advancements, the reader may consult \cite{guedj2019primer,alquier2024user}. Initially proposed in \cite{STW,McA}, PAC-Bayesian bounds were designed to empirically quantify the prediction risk for Bayesian estimators. However, as discussed in \cite{catoni2004statistical,catonibook}, this approach also proves useful for deriving non-asymptotic bounds.

In Theorem \ref{thm_main_2}, we establish a relationship between the integrated prediction risk of our approach and the minimal achievable risk \( R^* \). The boundedness assumption is not too crucial to our technical proofs,  \cite{alquier2024user} suggests that alternative methods might allow for its relaxation. In Theorem \ref{thm_heavy_tailed_loss} below, we provide a result for unbounded loss. Although the excess risk error rate in Theorem \ref{thm_main_2} is slower by an order of \( n^{-1/2} \) compared to the results in Theorem \ref{thm_main1}, it is, to our knowledge, an entirely new finding. Furthermore, it requires only minimal assumptions, such as a bounded loss function and a bounded moment for the distribution of the covariate \(X\).

We will now obtain results for the mean estimator as a direct consequence of Theorem \ref{thm_main_2}, utilizing Jensen's inequality and the convexity of the loss. Therefore, the proof is not included.

\begin{cor}
	Assume Theorem \ref{thm_main_2} holds true, then we have that
	\begin{equation*}
		\mathbb{E}\, R(\hat\theta) - R^*
		\leq 
		\mathcal{C}_1
		\tilde{\xi}
		,
	\end{equation*}	
	and	 with probability at least $ 1-\varepsilon, \varepsilon\in (0,1) $ that
	\begin{equation*}
		R(\hat\theta) - R^*
		\leq 
		\mathcal{C}'_1
		[ \tilde{\xi}+	n^{-1/2} \log( 2 / \varepsilon) ]
		,
	\end{equation*}	
	for some constant $ \mathcal{C}_1 , \mathcal{C}_1' > 0 $ depending only on $ C,C_1,C_{\rm x} , C_L $.
\end{cor}

We now present a result concerning randomized estimators where \( \theta \) is sampled from \( \hat{\rho}_{\lambda} \). Here, it should be noted that `\textit{with probability at least}' in Proposition \ref{propo_slow} below means the probability evaluated with respect to the distribution $ \mathbf{P}^{\otimes n} $ of the data and the conditional Gibbs posterior distribution $ \hat{\rho}_{\lambda} $.

\begin{prop}
	\label{propo_slow}
	Assume that Assumption \ref{assume_bounded_loss} and \ref{assume_X_bounded} are satisfied.  Take $\lambda= \sqrt{n} $, $ \varsigma = ( C_L C_{\rm x} n\sqrt{d})^{-1} $.  Then for all $ \theta^* $ such that $  \| \theta^*\|_1 \leq C_1 - 2d\varsigma $, we have with probability at least $ 1-\varepsilon, \varepsilon\in (0,1) $ that
	\begin{equation*} 
		R(\theta) - R^*
		\leq 
		\mathcal{C}''_1
		[ \tilde{\xi} +	n^{-1/2} \log( 2 / \varepsilon) ]
		,
	\end{equation*}
	for some constant $ \mathcal{C}''_1 > 0 $ depending only on $ C,C_1,C_{\rm x}, C_L $.
\end{prop}

\subsection{Fast rates}

Theorem \ref{thm_main_2} provides an oracle inequality that connects the integrated prediction risk of our method to the minimum achievable risk. However, these bounds can be improved by introducing additional assumptions.

Put $ C_{_{K,C}} : = \max(2 C_L K,C) $.

\begin{theorem}
	\label{thm_main1}
	Assume that Assumption \ref{assume_bounded_loss} \ref{assume_X_bounded},\ref{assume_Lipschitz} and \ref{assum_bernstein} are satisfied. Take $\lambda= n/C_{_{K,C}}  $, $ \varsigma = ( C_L C_{\rm x} n\sqrt{d})^{-1} $.  Then for all $ \theta^* $ such that $  \| \theta^*\|_1 \leq C_1 - 2d\varsigma $ we have:
	\begin{equation*}
		\mathbb{E} \mathbb{E}_{\theta\sim\hat{\rho}_{\lambda}} [R(\theta) ]- R^*
		\leq 
		\mathcal{C}_2
s^* \log ( n\sqrt{d} / s^*)
/ n
	\end{equation*}
	and	 with probability at least $ 1-\varepsilon, \varepsilon\in (0,1) $ that
	\begin{equation*}
		\mathbb{E}_{\theta\sim\hat{\rho}_{\lambda}} [R(\theta) ] - R^*
		\leq 
		\mathcal{C}_2'
		\frac{s^* \log ( n\sqrt{d} / s^*)  +	\log( 2 / \varepsilon)
			}{ n }
		,
	\end{equation*}	
	for some constant $ \mathcal{C}_2 , \mathcal{C}_2' > 0 $ depending only on $ K, C,C_1,C_{\rm x},C_L $.
\end{theorem}

In contrast to Theorem \ref{thm_main_2}, Theorem \ref{thm_main1} offers a bound that scales more rapidly, at \( 1/n \) rather than \( 1/\sqrt{n} \). Both results are presented in expectation and with high probability, facilitating a comparison between the out-of-sample error of our method and the optimal prediction error \( R^* \).

\begin{rmk}
It is worth noting that risk bounds for generalized linear models (GLMs) have been established for penalization methods in terms of Kullback-Leibler divergence in \cite{abramovich2016model} and \cite{abramovich2018high}. During the preparation of this manuscript, a preprint \cite{zilberman2024high} appeared, focusing on penalization methods for count responses. This paper builds upon the results from \cite{abramovich2018high} for specific regression models, such as Poisson and negative binomial regression, and derives explicit prediction error bounds. Additionally, it introduces algorithms for efficiently computing the penalized method. Our fast rate of \( s^* \log (d/s^*)/n \), in Theorem \ref{thm_main1}, aligns with the results from these works \cite{abramovich2016model,abramovich2018high,zilberman2024high}.
\end{rmk}

As a direct consequence of Theorem \ref{thm_main1} by using Jensen's inequality, the results for the mean estimator is given in the following corollary. Therefore, the proof is not included.

\begin{cor}
	Assume Theorem \ref{thm_main1} holds true, then we have that
	\begin{equation*}
		\mathbb{E} [R(\hat\theta) ]- R^*
		\leq 
		\mathcal{C}_2
s^* \log ( n\sqrt{d} / s^*)
/ n
	\end{equation*}
	and	 with probability at least $ 1-\varepsilon, \varepsilon\in (0,1) $ that
	\begin{equation*}
		R(\hat\theta) - R^*
		\leq 
		\mathcal{C}_2'
		\frac{s^* \log ( n\sqrt{d} / s^*)  +	\log( 2 / \varepsilon)
}{ n }
		,
	\end{equation*}	
	for some constant $ \mathcal{C}_2 , \mathcal{C}_2' > 0 $ depending only on $ K, C,C_1,C_{\rm x} $.
\end{cor}

Under Assumption \ref{assum_bernstein}, we are able to establish a fast excess risk bound. This assumption not only facilitates a fast rate of the excess risk but also allows us to gain deeper insights into the relationship between our predictor and the optimal predictor \( \theta^* \).

\begin{assume}
	\label{assum_eigen}
	Assume that there is a constant $ \kappa >0 $ such that, 
	\begin{equation*}
	\kappa^2	\|  \theta - \theta^* \|^2
	\leq
	\mathbb{E} \{ 
	|  X^\top (\theta - \theta^*) |^2 \}
	.
	\end{equation*}
\end{assume}

Assumption \ref{assum_eigen} is satisfied when the smallest eigenvalue of \( \mathbb{E} (XX^\top) \) is bounded below by a positive constant, which is similar to the sparse eigenvalue condition often employed in the high-dimensional regression literature \cite{buhlmann_vandegeer,abramovich2018high}. While this assumption is not critical for our results, it is necessary for deriving the error bounds for the predictor.

\begin{prop}
	\label{propo_fastrate}
	Assume that Theorem \ref{thm_main1} holds, we deduce that
	\begin{equation*}
		\mathbb{E}\, \mathbb{E}_{\theta\sim\hat{\rho}_{\lambda}}  \{ |  X^\top (\theta - \theta^*) |^2 \}
		\leq 
		C''_2
s^* \log ( n\sqrt{d} / s^*)
/ n
	\end{equation*}
	and further assume that Assumption \ref{assum_eigen} satisfied, then
	\begin{equation*}
		\mathbb{E}\, \mathbb{E}_{\theta\sim\hat{\rho}_{\lambda}}  
		\|  \theta - \theta^* \|^2
		\leq 
		\frac{C'''_2 }{ \kappa^2} 
	s^* \log ( n\sqrt{d} / s^*)
	/ n
	,
	\end{equation*}
	for some constant $ C''_2 > 0 $ depending only on $ K, C,C_1,C_{\rm x} $ and  $ C'''_2 > 0 $ depending only on $ \kappa, K, C,C_1,C_{\rm x} $.
\end{prop}

\begin{rmk}
Proposition \ref{propo_fastrate} provides non-asymptotic bounds on the mean-square error of our predictors. This rate aligns with the findings from \cite{zilberman2024high} and \cite{abramovich2018high} (see Lemma 1 in \cite{abramovich2016model}) and is minimax-optimal as established in Theorem 2 of \cite{abramovich2016model}.
\end{rmk}

\subsection{Fast rate with heavy-tailed loss}

The boundedness assumption can be relaxed, as done in \cite{alquier2024user}. Here, we explore a scenario where the loss is heavy-tailed, satisfying Assumption \ref{assum_heavy_tailed}, and establish a fast rate result under this condition.

\begin{theorem}
	\label{thm_heavy_tailed_loss}
	Assume that Assumption \ref{assume_X_bounded},\ref{assume_Lipschitz} and \ref{assum_heavy_tailed} are satisfied. Take $\lambda= n/(H_1 +2H_2)  $, $ \varsigma = ( C_L C_{\rm x} n\sqrt{d})^{-1} $.  Then for all $ \theta^* $ such that $  \| \theta^*\|_1 \leq C_1 - 2d\varsigma $ we have with probability at least $ 1-\varepsilon, \varepsilon\in (0,1) $ that
	\begin{equation*}
	\mathbb{E}_{\theta\sim\hat{\rho}_{\lambda}} [R(\theta) ] - R^*
	\leq 
	\mathcal{C}_3
	\frac{s^* \log ( n\sqrt{d} / s^*)  +	\log( 2 / \varepsilon)
	}{ n }
	,
	\end{equation*}	
	for some universal constant $ \mathcal{C}_3  > 0 $ depending only on $ C_1,C_{\rm x},C_L,H_1,H_2 $.
\end{theorem}

The prediction bounds derived in Theorems \ref{thm_main_2}, \ref{thm_main1}, and \ref{thm_heavy_tailed_loss} are novel contributions, with an explicit dependence on $s^* $, highlighting the adaptability of our approach in sparse settings. Importantly, the results from these theorems showcase adaptive properties, showing that the method's performance does not require knowledge of $s^* $, the sparsity level of $\theta^* $. When the true sparsity $s^*$ is particularly small, the prediction error nearly matches the optimal error, $R^*$, even with a limited sample size. This outcome is known as an ``oracle inequality", meaning that the estimator performs almost as well as if the sparsity level were known beforehand through an oracle.

In this section, we presented the selected values for the tuning parameters \( \lambda \) and \( \varsigma \) in our proposed method, which provide different theoretical prediction error rates. Although these values serve as a useful starting point, they may not be optimal in practical applications. Users can, for instance, apply cross-validation to fine-tune these parameters for specific tasks. Still, the theoretical values discussed here offer a valuable reference for assessing the appropriate scale of tuning parameters in real-world contexts.

\begin{table}[!h]
	\centering
	\caption{Simulation results with Poisson model. $ n = 50, p = 100$}
	\begin{tabular}{ l  ccc }
		\hline \hline
		error  & LMC & MALA & Lasso 
		\\
		\hline
		\multicolumn{4}{c}{	$ s^* = 5 $}
		\\
		mse$ \times 10 $
		& 0.035 (0.016) & 0.042 (0.016) & 0.034 (0.016)
		\\
		nsp
		& 0.318 (0.191) &  0.322 (0.185) & 0.318 (0.192)
		\\	
		mde	
		& 0.332 (0.247) & 0.344 (0.225) & 0.331 (0.250)
		\\
		\hline
		\multicolumn{4}{c}{	$ s^* = 5 $, noisy}
		\\
		mse $ \times 10 $
		& 0.065 (0.047) & 0.073 (0.047) & 0.065 (0.047)
		\\
		nsp
		& 0.449 (0.309) & 0.402 (0.293) & 0.452 (0.314)
		\\	
		mde	
		& 1.424 (1.412) & 1.248 (1.114) & 1.449 (1.453)
		\\
		\hline
		\multicolumn{4}{c}{	$ s^* = 25 $}
		\\
		mse$ \times 10 $
		& 0.181 (0.045) & 0.189 (0.047) & 0.180 (0.045)
		\\
		nsp
		& 0.095 (0.196) & 0.072 (0.171) & 0.088 (0.207)
		\\	
		mde	
		& 0.406 (0.919) & 0.377 (0.791) & 0.395 (0.995)
		\\
		\hline
		\multicolumn{4}{c}{	$ s^* = 25 $, noisy}
		\\
		mse$ \times 10 $
		& 0.242 (0.060) & 0.249 (0.063) & 0.240 (0.060)
		\\
		nsp
		& 0.152 (0.273) & 0.117 (0.232) & 0.168 (0.301)
		\\	
		mde	
		& 1.559 (3.404) & 1.739 (5.041) & 2.193 (6.076)
		\\
		\hline
		\hline
	\end{tabular}
	\label{tb_posson_small}
	
	\centering
	\caption{Simulation results with Poisson model. $ p = 300, n=100$}
	\begin{tabular}{ l  ccc }
		\hline \hline
		error  & LMC & MALA & Lasso 
		\\
		\hline
		\multicolumn{4}{c}{	$ s^* = 5 $}
		\\
		mse$ \times 10^3 $
		& 0.728 (0.351) & 0.852 (0.352) & 0.724 (0.351)
		\\
		nsp
		& 0.331 (0.140) & 0.341 (0.137) & 0.331 (0.141)
		\\	
		mde	
		& 0.333 (0.159) & 0.353 (0.155) & 0.334 (0.161)
		\\
		\hline
		\multicolumn{4}{c}{	$ s^* = 5 $, noisy}
		\\
		mse$ \times 10^2 $
		& 0.183 (0.110) & 0.196 (0.111) & 0.183 (0.110)
		\\
		nsp
		& 0.460 (0.324) & 0.448 (0.324) & 0.462 (0.325)
		\\	
		mde	
		& 1.587 (1.363) & 1.573 (1.313) & 1.600 (1.382)
		\\
		\hline
		\multicolumn{4}{c}{	$ s^* = 25 $}
		\\
		mse$ \times 10^2 $
		& 0.449 (0.137) & 0.462 (0.136) & 0.449 (0.137)
		\\
		nsp
		& 0.050 (0.129) & 0.052 (0.116) & 0.049 (0.130)
		\\	
		mde	
		& 0.237 (0.504) & 0.270 (0.459) & 0.236 (0.510)
		\\
		\hline
		\multicolumn{4}{c}{	$ s^* = 25 $, noisy}
		\\
		mse$ \times 10^2 $
		& 0.719 (0.185) & 0.732 (0.186) & 0.719 (0.185)
		\\
		nsp
		& 0.181 (0.302) & 0.163 (0.287) & 0.183 (0.305)
		\\	
		mde	
		& 1.872 (3.505) & 1.817 (3.325) & 1.899 (3.585)
		\\
		\hline
		\hline
	\end{tabular}
	\label{tb_posson_large}
\end{table}

\begin{table}[!h]
	\centering
	\caption{Simulation results with Negative Binomial model. $ p = 100, n= 50$,  dispersion parameter $ \alpha = 2 $}
	\begin{tabular}{ l  ccc }
		\hline \hline
		error  & LMC & MALA & Lasso 
		\\
		\hline
	\multicolumn{4}{c}{$ s^* = 5 $}	
		\\
		mse$ \times 10^2 $
		& 0.412 (0.176) & 0.481 (0.181) & 0.408 (0.176)
		\\
		nsp
		& 0.423 (0.231) & 0.415 (0.226) & 0.425 (0.235)
		\\	
		mde	
		& 0.539 (0.374) & 0.527 (0.328) & 0.543 (0.380)
		\\
		\hline
		\multicolumn{4}{c}{$ s^* = 5 $, noisy }
		\\
		mse$ \times 10^2 $
		& 0.814 (0.578) & 0.899 (0.593) & 0.810 (0.580)
		\\
		nsp
		& 0.423 (0.345) & 0.370 (0.325) & 0.426 (0.351)
		\\	
		mde	
		& 1.510 (1.697) & 1.237 (1.167) & 1.547 (1.772)
		\\
		\hline
	\multicolumn{4}{c}{	$ s^* = 25 $}
		\\
		mse$ \times 10 $
		& 0.203 (0.043) & 0.209 (0.045) & 0.202 (0.043)
		\\
		nsp
		& 0.205 (0.287) & 0.155 (0.239) & 0.206 (0.292)
		\\	
		mde	
		& 0.973 (1.590) & 0.821 (1.319) & 0.994 (1.683)
		\\
		\hline
	\multicolumn{4}{c}{	$ s^* = 25 $, noisy }
		\\
		mse$ \times 10 $
		& 0.249 (0.063) & 0.255 (0.065) & 0.248 (0.063)
		\\
		nsp
		& 0.236 (0.337) & 0.177 (0.288) & 0.243 (0.342)
		\\	
		mde	
		& 2.602 (3.913) & 1.988 (3.007) & 2.708 (4.121)
		\\
		\hline
		\hline
	\end{tabular}
	\label{tb_negB_small}
	
	\centering
	\caption{Simulation results  with Negative Binomial model. $ p = 300, n=100$,  dispersion parameter $ \alpha = 2 $}
	\begin{tabular}{ l  ccc }
		\hline \hline
		error  & LMC & MALA & Lasso 
		\\
		\hline
	\multicolumn{4}{c}{	$ s^* = 5 $}
		\\
		mse$ \times 10^2 $
		& 0.112 (0.045) & 0.124 (0.044) & 0.111 (0.045)
		\\
		nsp
		& 0.456 (0.223) & 0.461 (0.226) & 0.457 (0.224)
		\\	
		mde	
		& 0.632 (0.435) & 0.647 (0.430) & 0.635 (0.440)
		\\
		\hline
	\multicolumn{4}{c}{	$ s^* = 5 $, noisy }
		\\
		mse$ \times 10^2 $
		& 0.190 (0.108) & 0.204 (0.108) & 0.190 (0.108)
		\\
		nsp
		& 0.478 (0.341) & 0.466 (0.340) & 0.480 (0.342)
		\\	
		mde	
		& 1.922 (1.705) & 1.905 (1.690) & 1.934 (1.718)
		\\
		\hline
	\multicolumn{4}{c}{	$ s^* = 25 $ }
		\\
		mse$ \times 10^2 $
		& 0.594 (0.163) & 0.607 (0.163) & 0.594 (0.163)
		\\
		nsp
		& 0.145 (0.267) & 0.137 (0.253) & 0.144 (0.268)
		\\	
		mde	
		& 0.971 (2.421) & 0.946 (2.217) & 0.979 (2.463)
		\\
		\hline
	\multicolumn{4}{c}{	$ s^* = 25 $, noisy }
		\\
		mse$ \times 10^2 $
		& 0.790 (0.167) & 0.799 (0.165) & 0.790 (0.167)
		\\
		nsp
		& 0.206 (0.339) & 0.192 (0.323) & 0.209 (0.341)
		\\	
		mde	
		& 2.827 (5.175) & 2.739 (5.023) & 2.887 (5.324)
		\\
		\hline
		\hline
	\end{tabular}
	\label{tb_negB_large}
\end{table}

\begin{table}[!h]
	\centering
	\caption{Simulation results  with Negative Binomial model. $ p = 100, n=50$, dispersion parameter $ \alpha = 20 $}
	\begin{tabular}{ l  ccc }
		\hline \hline
		error  & LMC & MALA & Lasso 
		\\
		\hline
	\multicolumn{4}{c}{	$ s^* = 5 $ }
		\\
		mse$ \times 10^2 $
		& 0.372 (0.182) & 0.444 (0.192) & 0.367 (0.182)
		\\
		nsp
		& 0.326 (0.200) & 0.336 (0.191) & 0.327 (0.202)
		\\	
		mde	
		& 0.354 (0.265) & 0.377 (0.256) & 0.354 (0.269)
		\\
		\hline
	\multicolumn{4}{c}{	$ s^* = 5 $, noisy }
		\\
		mse$ \times 10^2 $
		& 0.727 (0.472) & 0.800 (0.475) & 0.721 (0.474)
		\\
		nsp
		& 0.347 (0.327) & 0.301 (0.305) & 0.347 (0.331)
		\\	
		mde	
		& 1.276 (1.528) & 1.099 (1.223) & 1.296 (1.573)
		\\
		\hline
	\multicolumn{4}{c}{	$ s^* = 25 $ }
		\\
		mse$ \times 10 $
		& 0.179 (0.047) & 0.187 (0.045) & 0.179 (0.047)
		\\
		nsp
		& 0.135 (0.254) & 0.105 (0.212) & 0.133 (0.258)
		\\	
		mde	
		& 0.644 (1.331) & 0.548 (0.968) & 0.652 (1.374)
		\\
		\hline
	\multicolumn{4}{c}{	$ s^* = 25 $, noisy }
		\\
		mse$ \times 10 $
		& 0.238 (0.062) & 0.245 (0.067) & 0.238 (0.062)
		\\
		nsp
		& 0.220 (0.341) & 0.187 (0.313) & 0.224 (0.351)
		\\	
		mde	
		& 2.545 (6.131) & 2.261 (5.767) & 2.890 (7.692)
		\\
		\hline
		\hline
	\end{tabular}
	\label{tb_negB_samll_withtheta}
	
	\centering
	\caption{Simulation results  with Negative Binomial model. $ p = 300, n=100$, dispersion parameter $ \alpha = 20 $}
	\begin{tabular}{ l  ccc }
		\hline \hline
		error  & LMC & MALA & Lasso 
		\\
		\hline
	\multicolumn{4}{c}{	$ s^* = 5 $ }
		\\
		mse$ \times 10^2 $
		& 0.084 (0.035) & 0.096 (0.036) & 0.083 (0.035)
		\\
		nsp
		& 0.333 (0.164) &  0.348 (0.165) & 0.334 (0.165)
		\\	
		mde	
		& 0.342 (0.186) & 0.369 (0.186) & 0.343 (0.187)
		\\
		\hline
	\multicolumn{4}{c}{	$ s^* = 5 $, noisy }
		\\
		mse$ \times 10^2 $
		& 0.186 (0.109) & 0.199 (0.109) & 0.185 (0.109)
		\\
		nsp
		& 0.458 (0.323) & 0.450 (0.323) & 0.459 (0.324)
		\\	
		mde	
		& 1.556 (1.352) & 1.561 (1.339) & 1.564 (1.365)
		\\
		\hline
	\multicolumn{4}{c}{	$ s^* = 25 $ }
		\\
		mse$ \times 10^2 $
		& 0.491 (0.133) & 0.502 (0.133) & 0.490 (0.133)
		\\
		nsp
		& 0.078 (0.174) & 0.076 (0.155) & 0.078 (0.176)
		\\	
		mde	
		& 0.374 (0.803) & 0.396 (0.732) & 0.374 (0.811)
		\\
		\hline
	\multicolumn{4}{c}{	$ s^* = 25 $, noisy }
		\\
		mse$ \times 10^2 $
		& 0.743 (0.167) & 0.754 (0.166) &  0.744 (0.167)
		\\
		nsp
		& 0.250 (0.356) & 0.234 (0.348) &  0.252 (0.358)
		\\	
		mde	
		& 3.871 (7.037) & 3.708 (6.550) & 3.950 (7.276)
		\\
		\hline
		\hline
	\end{tabular}
	\label{tb_negB_large_withtheta}
\end{table}

\section{Numerical studies}
\label{sc_numberical}
\subsection{Implementation}

Up to a constant, the gradient of the log of the posterior in \eqref{eq_Gibbs_poste} is as
\begin{equation*}
\nabla \log (\hat{\rho}_{\lambda} (\theta) )
=
- \lambda \nabla r_n(\theta) + \nabla \log (\pi (\theta))
\end{equation*}
and $ \nabla r_n(\theta) $ and $ \nabla \log (\pi (\theta)) $ can be derived explicitly. This enables the use of the Langevin Monte Carlo (LMC) method, a gradient-based sampling technique, to draw samples from the Gibbs posterior in \eqref{eq_Gibbs_poste}. LMC has been recognized as an effective method for sampling in high-dimensional settings, as noted in \cite{durmus2019high} and \cite{dalalyan2017theoretical}.

We employ both the constant step-size unadjusted Langevin Monte Carlo (LMC) and the Metropolis-adjusted Langevin algorithm (MALA) to approximate our method. While LMC is computationally efficient, it only approximately targets the intended distribution. In contrast, MALA, which incorporates an additional Metropolis–Hastings correction step, guarantees convergence to the desired distribution, albeit at the cost of increased computational complexity.

We compare our proposed methods, referred to as LMC and MALA, with the state-of-the-art Lasso method, available from the \texttt{R} package \texttt{glmnet} \citep{glmnet}. 

\subsection{Simulation studies}
\subsubsection{Simulation setup}
To evaluate the performance of our proposed methods, we consider two distinct scenarios for modeling count data:
\begin{itemize}
	\item Poisson log-linear regression model:
$ \quad
	Y_i \sim Pois(\mu_i), \quad \log \mu_i = X_i^\top \theta^*
	;
$
	
	\item Negative Binomial (NB) regression:
	\begin{align*}
	Y_i \sim NB(\alpha , \frac{\alpha}{\alpha + \mu_i} ), \quad \log \mu_i = X_i^\top \theta^*
	\end{align*} 
	with some given $ \alpha >0 $.
\end{itemize}
The Poisson regression model stipulates that both the mean and variance of \( Y_i \) (given \( X_i \)) are equal to \( \mu_i \). However, count data may exhibit overdispersion, characterized by the condition \( \text{Var}(Y_i) > \mathbb{E}[Y_i] \). In such cases, Negative Binomial regression is a widely used method for modeling overdispersed count data. In the above NB regression model, one has that $ \mathbb{E}[Y_i] = \mu_i $ and $ \text{Var}(Y_i) = \mu_i + \mu_i^2/\alpha > \mu_i $.

We examine two different scenarios regarding the dimensions: \( p = 100, n = 50 \) for a smaller scale and \( p = 300, n = 100 \) for a larger scale. In each case, the entries of the design matrix \( X \) are i.i.d. simulated from \( \mathcal{N}(0,1) \). For every configuration, we assess two levels of sparsity \( s^* \), which represents the number of non-zero elements in the parameter vector \( \theta^* \). Specifically, we analyze \( s^* = 5 \) and \( s^* = 25 \). In all scenarios, the non-zero elements of the true parameter vector \( \theta^* \) are independently sampled from a uniform distribution \( \mathcal{U}[-0.5, 0.5] \). Additionally, for the Negative Binomial regression model, we consider two distinct values of dispersion parameter \( \alpha \): \( \alpha = 2 \) and \( \alpha = 20 \). A ``noisy" model is also examined for each considered setting, defined as 
$$
 \log \mu_i = X_i^\top \theta^* + u_i, \text{ where } \,  u_i \sim \mathcal{N}(0,1) 
.
$$

To evaluate various methods, we utilize the following measures. Firstly, we consider the average squared estimation error to assess the performance of the predictor $ \widehat{\theta} $,
\begin{equation*}
{\rm mse} := d^{-1} \sum_{i=1}^d (\widehat{\theta} - \theta^* )^2
.
\end{equation*}
Next, we quantify prediction error using two distinct measures: the normalized squared prediction error and the empirical deviance error, defined as follows:
\begin{equation*}
{\rm nsp} 
:=
\frac{ n^{-1} \sum_{i=1}^n (Y_i - e^{X_i^\top \widehat{\theta} } )^2 }{ n^{-1} \sum_{i=1}^n (Y_i)^2 }
; \quad
{\rm mde} 
:=
n^{-1} \sum_{i=1}^n 
\left[
Y_i \log (Y_i/e^{X_i^\top \widehat{\theta} }) -Y_i + e^{X_i^\top \widehat{\theta} } 
\right]
.
\end{equation*}
Each simulation setup is independently repeated 100 times, and we present the average results along with their standard deviations. The findings are summarized in Tables \ref{tb_posson_small}, \ref{tb_posson_large}, \ref{tb_negB_small}, \ref{tb_negB_large}, \ref{tb_negB_samll_withtheta}, and \ref{tb_negB_large_withtheta}.

LMC and MALA are executed for 25,000 iterations, discarding the first 5,000 steps as burn-in. Both LMC and MALA are initialized using the Lasso estimate. For all scenarios, the tuning parameters are set to \( \lambda = 1 \) and \( \varsigma = 0.1 \). It is worth noting that a more effective approach would involve tuning these parameters using cross-validation, which could enhance the results. The Lasso method is implemented with default settings, utilizing 5-fold cross-validation to determine the optimal tuning parameter.

\subsubsection{Simulation results}

The simulation results indicate that our proposed methods hold up remarkably well against the Lasso method, demonstrating their competitive edge. While MALA may occasionally exhibit slightly higher estimation errors, it consistently delivers superior predictive performance compared to Lasso, particularly when dealing with small sample sizes. This advantage becomes even more pronounced when the data set is 'noisy,' where MALA shows a significant improvement in prediction error. These observations suggest that our methods are not only robust in handling smaller, less reliable data sets, but also offer clear benefits in complex modeling situations where noise is a major factor. Overall, the simulations confirm that our approach offers a reliable alternative to Lasso, especially in challenging scenarios.

Another important observation is that our proposed method performs well in both Poisson regression and Negative Binomial regression models. While all the estimation errors increase slightly when applied to the Negative Binomial regression model, this trend is consistent with what we see in the Lasso method. This behavior reinforces the idea that our method is flexible and adaptable to the underlying model structure of count data. Whether the data follows a Poisson or Negative Binomial distribution, our method maintains competitive performance, further showcasing its robustness across different statistical frameworks.

\subsection{An application: predicting the number of affairs}

The data set \texttt{affairs} in  the \texttt{R} package \texttt{COUNT} is a data set with 601 observations and 18 variables, including, naffairs (the number of affairs within the last year), and 17 covariates about children, marriage happiness, being religious, and age levels. The variable naffairs is considered as the response variable. 

For comparison, we randomly select 421 out of the 601 samples as training data, with the remaining 180 samples used as test data, resulting in a roughly 70/30 percent of the data split. The methods are executed on the training set, and their predictive accuracy is subsequently evaluated on the test set. This procedure is repeated 100 times with different random splits of training and testing data each time. Repeating and averaging the process across multiple iterations provides a more stable and precise assessment of the methods, helping to mitigate the impact of data variability and enhancing our understanding of their performance. The results are summarized in Table \ref{tb_affairs}.

\begin{table}[!h]
	\centering
	\caption{Means (and standard deviations) prediction errors for the  \texttt{affairs} data}
	\begin{tabular}{  l  ccc  }
		\hline \hline
		error & LMC & MALA & Lasso 
		\\ 
		\hline
nsp & 0.680 (0.024) & 0.659 (0.023) & 0.687 (0.029)
		\\
mde & 4.739 (0.729) & 4.694 (0.806) & 4.957 (0.769)
		\\
		\hline
		\hline
	\end{tabular}
	\label{tb_affairs}
\end{table}

Based on the results in Table \ref{tb_affairs}, it is evident that our proposed method, calculated using the MALA algorithm, outperforms the others in terms of prediction accuracy. The LMC method ranks second, while Lasso trails in third place. These findings highlight the strong performance of our approach—not just theoretically and in simulations, but also when applied to real data.

\section{Conclusion}
\label{sc_conlsution}

This paper presents a novel probabilistic machine learning framework specifically tailored for predicting high-dimensional count data. We introduce a pseudo-Bayesian approach that utilizes a scaled Student prior to promote sparsity and incorporates an exponential weight aggregation technique. A significant contribution of our study is the development of a new risk measure (loss function) designed for count data prediction, accompanied by theoretical guarantees for prediction risk established through PAC-Bayesian bounds. Our findings include non-asymptotic oracle inequalities that demonstrate the potential for rate-optimal prediction error without prior knowledge of sparsity. We derive both slow and fast rates under various conditions related to the loss function.

Additionally, we propose an efficient method for sampling from the pseudo-posterior and calculating the mean estimator using the Langevin Monte Carlo method. The strong performance of our approach, as evidenced by simulations and applications to real data, highlights its advantages over the Lasso method in a variety of contexts.

In this study, we did not address zero-inflated count response data. However, in certain real-world scenarios, the data may contain more zeros than what the Poisson model predicts, resulting in overdispersion. In such cases, the Zero-Inflated Poisson Model could be a suitable alternative. Therefore, extending our approach to account for this would be a potential direction for future research. Additionally, we did not explore variable selection, which is another important aspect. Future work should consider this to improve the practical applicability of our method.

\subsubsection*{Acknowledgments}
The author acknowledges support from the Centre for Geophysical Forecasting, Norwegian Research Council grant no. 309960, at NTNU. 

\subsubsection*{Conflicts of interest/Competing interests}
The author declares no potential conflict of interests.

\appendix
\section{Proofs}
\label{sc_proofs}

Let \( \mathcal{P}(\Theta) \) represent the collection of all probability measures defined on \(\Theta\).

\subsection{Proof for slow rate}

\begin{proof}[\bf Proof for Theorem \ref{thm_main_2}]
	\text{}
	\\
	\noindent \textit{Step 1, obtaining general PAC-Bayes bounds:}
	
Let $\theta \in \Theta$ be fixed, with $\ell_{i}(\theta) = (Y_i - \exp(X_i^\top \theta))^2$, and define $W_i = \ell_i(\theta^*) - \ell_i(\theta)$. It follows that $\mathbb{E} W_i = R(\theta^*) - R(\theta)$. By using Hoeffding's Lemma \ref{lemma:hoeffding} for $ W_i $, under Assumption \ref{assume_bounded_loss}, we obtain for any $t>0$ that
	$
	\mathbb{E}  {\rm e}^{ t n 
		[R(\theta) - R(\theta^*) -r(\theta) + r(\theta^*)] }  
	\leq 
	{\rm e}^{\frac{n t^2 C^2}{ 8}}.
	$
	Set $t = \lambda / n$, leading to the following expression:
	\begin{equation*}
	\mathbb{E}  {\rm e}^{ \lambda 	[R(\theta) - R(\theta^*) -r(\theta) + r(\theta^*)]  } 
	\leq 
	{\rm e}^{\frac{\lambda^2 C^2}{ 8 n}}.
	\end{equation*}
	By integrating this bound with respect to $\pi$ and applying Fubini's theorem, we arrive at:
	\begin{equation}
	\label{eq_bound_for_randomized_slow}
	\mathbb{E}  \mathbb{E}_{\theta\sim\pi} \left[ {\rm e}^{ \lambda 	
		[R(\theta) - R(\theta^*) -r(\theta) + r(\theta^*)] } \right] 
	\leq 
	{\rm e}^{\frac{\lambda^2 C^2}{ 8 n}}
	.
	\end{equation}
	Next, we use Lemma~\ref{lemma:dv} to obtain:
	\begin{equation*}
	\mathbb{E} \left[ {\rm e}^{ \sup_{\rho\in\mathcal{P}(\Theta)} \lambda \mathbb{E}_{\theta\sim\rho}
		[R(\theta) - R(\theta^*) -r(\theta) + r(\theta^*)] -	\mathcal{K}(\rho\|\pi) } \right] 
	\leq 
	{\rm e}^{\frac{\lambda^2 C^2}{ 8 n}}.
	\end{equation*} 
	Reorganizing the terms yields:
	\begin{equation}
	\label{eq_to_provein_probabiliy}
	\mathbb{E} \left[ {\rm e}^{ \sup_{\rho\in\mathcal{P}(\Theta)} \lambda \mathbb{E}_{\theta\sim\rho}
		[R(\theta) - R(\theta^*) -r(\theta) + r(\theta^*)] -	\mathcal{K}(\rho\|\pi)-\frac{\lambda^2 C^2}{ 8 n} } \right] 
	\leq 
	1
	.
	\end{equation}
	Utilizing Chernoff's bound, we obtain:
	\begin{equation*}
	\mathbb{P} \left[ \sup_{\rho\in\mathcal{P}(\Theta)} \lambda \mathbb{E}_{\theta\sim\rho}
	[R(\theta) - R(\theta^*) -r(\theta) + r(\theta^*)]
	-	\mathcal{K}(\rho\|\pi)-\frac{\lambda^2 C^2}{ 8 n}  > \log\frac{1}{\varepsilon} \right] 
	\leq 
	\varepsilon.
	\end{equation*} 
	By considering the complement and reorganizing the terms, we find that the probability is at least  $ 1- \varepsilon $
	\begin{equation*}
	\mathbb{E}_{\theta\sim\rho}[ R(\theta) - R(\theta^*)] 
	\leq
	\mathbb{E}_{\theta\sim\rho}[ r(\theta) -  r(\theta^*)] +  \frac{\lambda C^2}{8n} + \frac{	\mathcal{K}(\rho\|\pi) + \log\frac{1}{\varepsilon}}{\lambda} 
	.
	\end{equation*}
	By utilizing Lemma \ref{lemma:dv} again, we find that with probability at least $ 1- \varepsilon $
	\begin{equation}
	\label{eq_pacb_slow_1}
	\mathbb{E}_{\theta \sim \hat{\rho}_\lambda}[ R(\theta) - R(\theta^*)] 
	\leq
	\inf_{\rho\in\mathcal{P}(\Theta)}  
	\left\{ 
	\mathbb{E}_{\theta\sim\rho}[ r(\theta) -  r(\theta^*)] +  \frac{\lambda C^2}{8n} + \frac{	\mathcal{K}(\rho\|\pi) + \log\frac{1}{\varepsilon}}{\lambda} 
	\right\} 
	.
	\end{equation}	
	By applying the same argument to \( -W_i \) instead of \( W_i \), one obtains   with probability at least $ 1- \varepsilon $ that
	\begin{equation}
	\label{eq_pacb_slow_2}
	\mathbb{E}_{\theta\sim\rho}[ r(\theta) -  r(\theta^*)]
	\leq
	\mathbb{E}_{\theta\sim\rho}[ R(\theta) - R(\theta^*)] 
	+  \frac{\lambda C^2}{8n} + \frac{	\mathcal{K}(\rho\|\pi) + \log\frac{1}{\varepsilon}}{\lambda} 
	.
	\end{equation}
	Merging \eqref{eq_pacb_slow_1} and \eqref{eq_pacb_slow_2} yields, with a probability of at least $ 1- 2\varepsilon $, that
	\begin{equation}
	\label{eq_pacb_slow_3}
	\mathbb{E}_{\theta \sim \hat{\rho}_\lambda}[ R(\theta) - R(\theta^*)] 
	\leq
	\inf_{\rho\in\mathcal{P}(\Theta)}  \left\{ 
	\mathbb{E}_{\theta\sim\rho}
	[ R(\theta) - R(\theta^*)] 
	+  2\frac{\lambda C^2}{8n} 
	+ 2 \frac{	\mathcal{K} (\rho\|\pi) + \log\frac{1}{\varepsilon}}{\lambda} 
	\right\} 
	.
	\end{equation}
	
	To obtain bounds in expectation, from \eqref{eq_to_provein_probabiliy}, in particular for $\rho=\hat{\rho}_\lambda$, we use Jensen's inequality and rearranging terms:
	\begin{equation*}
	\lambda \left\{ \mathbb{E} \mathbb{E}_{\theta\sim\hat{\rho}_{\lambda}} [R(\theta) ]-R(\theta^*) \right\}
	\leq 
	\mathbb{E} \left[ \lambda [ \mathbb{E}_{\theta\sim\hat{\rho}_{\lambda} } [r(\theta) ]-r(\theta^*)] + 	
	\mathcal{K}(\hat{\rho}_{\lambda}\|\pi)
	+
	\frac{\lambda^2 C^2}{ 8 n}
	\right].
	\end{equation*}
	For $ \lambda >0 $ and note that $\hat{\rho}_\lambda$ minimizes the quantity in the expectation in the right-hand side (Lemma \ref{lemma:dv}), this can be rewritten:
	\begin{align}
	\mathbb{E} \mathbb{E}_{\theta\sim\hat{\rho}_{\lambda}} [R(\theta) ]-R(\theta^*)
	& \leq 
	\mathbb{E}  \inf_{\rho\in\mathcal{P}(\Theta)}  
	\left[   \mathbb{E}_{\theta\sim \rho} [r(\theta) ]-r(\theta^*) + 	
	\frac{\mathcal{K}(\rho\|\pi)}{\lambda}
	+
	\frac{\lambda C^2}{ 8 n}
	\right] \nonumber
	\\
	&\leq 
	\inf_{\rho\in\mathcal{P}(\Theta)}  
	\left[   \mathbb{E}_{\theta\sim \rho} 
	\mathbb{E}  [r(\theta) -r(\theta^*)] + 	
	\frac{\mathcal{K}(\rho\|\pi)}{\lambda}
	+
	\frac{\lambda C^2}{ 8 n}
	\right] \nonumber
	\\
	&\leq 
	\inf_{\rho\in\mathcal{P}(\Theta)}  
	\left[   \mathbb{E}_{\theta\sim \rho} 
	[R(\theta)] - R ( \theta^*) + 	
	\frac{\mathcal{K}(\rho\|\pi)}{\lambda}
	+
	\frac{\lambda C^2}{ 8 n}
	\right]
	\label{eq_bound_in_EXPECT_slow}
	.
	\end{align}

	\noindent \textit{Step 2, deriving specific rates:}
	
	In order to derive a particular rate, we next limit the infimum in \eqref{eq_pacb_slow_3} by setting \(\rho := p_0\), specified in \eqref{eq_specific_distribution}.
	
	Under Assumption \ref{assume_X_bounded} and the Lipschitz condition in Assumption \ref{assume_Lipschitz}, one has that
	$
	R (\theta) - R (\theta^*) 
	\leq
	C_L	\mathbb{E}	\|X_1\| \|\theta - \theta^* \|
	.
	$
	Now, using Lemma \ref{lema_bound_prior_arnak}, we have that
	\begin{align}
	\int [ R (\theta) - R (\theta^*) ] p_0 ({\rm d} \theta)
&	\leq
	C_L	\int \mathbb{E}	\|X_1\| \| \theta- \theta^* \| p_0( {\rm d} \theta) 
	\nonumber
	\\
&	\leq
	C_L		C_{\rm x}
	\left( \int \| \theta- \theta^* \|^2 p_0( {\rm d} \theta ) \right)^{1/2} 
	\leq
	C_L		C_{\rm x}
	\sqrt{	4d\varsigma^2 }
	\label{eq_bound_1_1}
	\end{align}
	and
	\begin{align}
	\mathcal{K}(p_0 \| \pi)
	\leq
	4 s^* \log \left(\frac{C_1 }{\varsigma s^*}\right)
	+
	\log(2)
	\label{eq_bound_1_2}
	.
	\end{align}
	Plug-in the bounds in \eqref{eq_bound_1_1} and \eqref{eq_bound_1_2} into inequality \eqref{eq_pacb_slow_3} and take $ \lambda = \sqrt{n} $, one gets with probability at least $ 1- 2\varepsilon $ that
	\begin{equation*}
	\mathbb{E}_{\theta\sim \hat{\rho}_{\lambda}}
	[ R (\theta) ] - R^*
	\leq
	\inf_{\varsigma \in (0,C_1/2d)} 
	\!
	\left\{ 
	C_L	C_{\rm x} 2 \varsigma \sqrt{ d }
	+  
	\frac{ C^2}{4\sqrt{n}} 
	+ 2 \frac{ 4 s^* \log \left(\frac{C_1 }{\varsigma s^*}\right)
		+
		\log(2) + \log\frac{1}{\varepsilon}}{\sqrt{n} } 
	\right\} 
	,
	\end{equation*}
	and the choice $ \varsigma = ( C_L	 C_{\rm x} n\sqrt{d})^{-1} $ leads to
	\begin{align*}
	\mathbb{E}_{\theta\sim \hat{\rho}_{\lambda}}
	[ R (\theta) ] - R^*
	& \leq 
	\frac{2}{n} +  \frac{ C^2}{4\sqrt{n}} 
	+ 
	2 \frac{ 4 s^* \log \left(\frac{C_1 C_L	 C_{\rm x} n\sqrt{d} }{ s^*}\right)
		+
		\log(2) + \log\frac{1}{\varepsilon}}{\sqrt{n} } 
	\\ 
	& \leq
	C_{_{C,C_1,L,X}}
	\frac{ s^* \log \left(\frac{ n\sqrt{d}}{ s^*}\right)  + \log\frac{1}{\varepsilon}
	}{\sqrt{n} }
	,
	\end{align*}
	for some constant $ C_{_{C,C_1,X,L}} > 0 $ depending only on $ C,C_1,C_L	,C_{\rm x} $.
	
	Now, plug-in the bounds in \eqref{eq_bound_1_1} and \eqref{eq_bound_1_2} into inequality \eqref{eq_bound_in_EXPECT_slow} and take $ \lambda = \sqrt{n} $, we obtain that
	\begin{equation*}
	\mathbb{E}\, 
	\mathbb{E}_{\theta\sim \hat{\rho}_{\lambda}}
	[ R (\theta) ] - R^*
	\leq
	\inf_{\varsigma \in (0,C_1/2d)} 
	\!
	\left\{ 
	C_L	C_{\rm x} 2 \varsigma \sqrt{	d }
	+  \frac{ C^2}{ 8 \sqrt{n}} 
	+  \frac{ 4 s^* \log \left(\frac{C_1 }{\varsigma s^*}\right)
		+	\log(2) }{\sqrt{n} } 
	\right\} 
	,
	\end{equation*}
	and the choice $ \varsigma = ( C_L	 C_{\rm x} n\sqrt{d})^{-1} $ leads to
	\begin{align*}
	\mathbb{E}\, 
	\mathbb{E}_{\theta\sim \hat{\rho}_{\lambda}}
	[ R (\theta) ] - R^*
	\leq 
	\frac{2}{n} +  \frac{ C^2}{8\sqrt{n}} 
	+ 
	\frac{ 4 s^* \log \left(\frac{C_1 C_{\rm x} C_L n\sqrt{d} }{ s^*}\right)
		+
		\log(2) }{\sqrt{n} } 
	\leq
	\mathcal{C}
	\frac{ s^* \log \left(\frac{ n\sqrt{d}}{ s^*}\right)  
	}{\sqrt{n} }
	,
	\end{align*}
	for some constant $\mathcal{C} > 0 $ depending only on $ C,C_1,C_{\rm x},C_L $.
	The proof is completed.
	
\end{proof}

\begin{proof}[\bf Proof of Proposition \ref{propo_slow}]
	From \eqref{eq_bound_for_randomized_slow}, for any $ \varepsilon \in (0,1) $, we have that	
	\begin{align*}
	\mathbb{E} \Biggl[ \int \exp \left\{ 
	\lambda [R( \theta )-R^* ] 
	-\lambda[r_n( \theta )-r_n^* ]
	- 
	\log \left[\frac{d\hat{\rho}_{\lambda}}{d \pi} (\theta)  \right]
	-
	\frac{\lambda^2 C^2 }{8n} 
	- 
	\log\frac{1}{\varepsilon}
	\right\}
	\hat{\rho}_{\lambda}(d \theta)
	\Biggr]
	\leq 
	\varepsilon
	.
	\end{align*}
	Using the fundamental inequality \(\exp(x) \geq \mathbf{1}_{\mathbb{R}_{+}}(x)\), it follows that with probability at most
	$\varepsilon $,
	\begin{align*}
	\lambda [R( \theta )-R^* ] 
	\geq
	\lambda[r_n( \theta )-r_n^* ]
	+
	\log \left[\frac{d\hat{\rho}_{\lambda}}{d \pi} (\theta)  \right]
	+
	\frac{\lambda^2 C^2 }{8n} 
	+
	\log\frac{1}{\varepsilon}
	,
	\end{align*}
	where the probability is evaluated with respect to the distribution $\mathbf
	P^{\otimes n}$ of the
	data {\it and} the conditional probability measure $\hat \rho_{\lambda} $.	Taking the complement, with probability at least $ 1-\varepsilon $, one has for $ \lambda>0 $ that
	\begin{align*}
	R( \theta )-R^* 
	\leq
	r_n( \theta )-r_n^* 
	+
	\frac{\log \left[\frac{d\hat{\rho}_{\lambda}}{d \pi} (\theta)  \right]
		+
		\frac{\lambda^2 C^2 }{8n} 
		+
		\log\frac{1}{\varepsilon}
	}{\lambda}
	.
	\end{align*}
	Observe that
	\begin{align*}
	\log\left(
	\frac{\mbox{d}\hat{\rho}_{\lambda}}{\mbox{d}\pi}(\theta)
	\right)
	= 
	\log
	\frac{\exp \left [-\lambda
		r_{n}(\theta)\right]
	}{ \int \exp\left[-\lambda r_{n}(\theta)\right]
		\mbox{d}\pi(\theta) }
	= 
	-\lambda r_{n}(\theta) - \log \int
	\exp\left[-\lambda r_{n}(\theta)\right] \mbox{d}\pi(\theta)
	,
	\end{align*}
	thus, we obtain with probability at least $ 1-\varepsilon $ that
	\begin{align*}
	R( \theta )-R^* 
	\leq
	- \frac{1}{\lambda} \log \int
	\exp\left[-\lambda r_{n}(\theta)\right] \mbox{d}\pi(\theta) - r_n^* 
	+
	\frac{	\frac{\lambda^2 C^2 }{8n} 
		+
		\log\frac{1}{\varepsilon}
	}{\lambda}
	.
	\end{align*}
	Now, using Lemma \ref{lemma:dv}, we get with probability at least $ 1-\varepsilon $ that
	\begin{align*}
	R( \theta )-R^* 
	\leq
	\inf_{\rho\in\mathcal{P}(\Theta)}  
	\left\{ 
	\mathbb{E}_{\theta\sim\rho}[ r(\theta) -  r(\theta^*)] +  \frac{\lambda C^2}{8n} + \frac{	\mathcal{K}(\rho\|\pi) + \log\frac{1}{\varepsilon}}{\lambda} 
	\right\} 
	.
	\end{align*}	
	Combining the above inequality wit \eqref{eq_pacb_slow_2}, we get with probability at least $ 1- 2\varepsilon $ that
	\begin{align}
	\label{eq_randomized_slow}
	R( \theta )-R^* 
	\leq
	\inf_{\rho\in\mathcal{P}(\Theta)}  \left\{ 
	\mathbb{E}_{\theta\sim\rho}
	[ R(\theta) - R(\theta^*)] 
	+  2\frac{\lambda C^2}{8n} 
	+ 2 \frac{	\mathcal{K} (\rho\|\pi) + \log\frac{1}{\varepsilon}}{\lambda} 
	\right\} 
	.
	\end{align}	
	We restrict the infimumn in \eqref{eq_randomized_slow} to $ \rho:= p_0 $ given in \eqref{eq_specific_distribution}.
	Using Lemma \ref{lema_bound_prior_arnak}, we can plug-in the bounds in \eqref{eq_bound_1_1} and \eqref{eq_bound_1_2} into inequality \eqref{eq_randomized_slow}, with $ \lambda = \sqrt{n} $, one gets that	
	\begin{equation*}
	R (\theta)  - R^*
	\leq
	\inf_{\tau \in (0,C_1/2d)} 
	\!
	\left\{ 
	C_L	C_{\rm x} 2 \varsigma \sqrt{	d }
	+  \frac{ C^2}{4\sqrt{n}} 
	+ 2 \frac{ 4 s^* \log \left(\frac{C_1 }{\varsigma s^*}\right)
		+
		\log(2) + \log\frac{1}{\varepsilon}}{\sqrt{n} } 
	\right\} 
	,
	\end{equation*}
	and the choice $ \varsigma = (C_L C_{\rm x} n\sqrt{d})^{-1} $ leads to
	\begin{align*}
	R (\theta) - R^*
	\leq 
	\frac{2}{n} +  \frac{ C^2}{4\sqrt{n}} 
	+ 
	2 \frac{ 4 s^* \log \left(\frac{ C_L C_1 C_{\rm x} n\sqrt{d} }{ s^*}\right)
		+
		\log(2) + \log\frac{1}{\varepsilon}}{\sqrt{n} } 
	\leq
	\mathcal{C}
	\frac{ s^* \log \left(\frac{ n\sqrt{d}}{ s^*}\right)  + \log\frac{1}{\varepsilon}
	}{\sqrt{n} }
	,
	\end{align*}
	for some constant $ \mathcal{C} > 0 $ depending only on $ C,C_1,C_{\rm x}, C_L $.
	
\end{proof}

\subsection{Proof for fast rate}

\begin{proof}[\bf Proof of Theorem \ref{thm_main1}]	
 From Assumption \ref{assume_Lipschitz} and \ref{assum_bernstein}, one obtains that
	\begin{align*}
	\mathbb{E} \{ (\ell_{i}(\theta) - \ell_{i}(\theta^*) )^2 \}
	\leq
	C_L K [ R (\theta) - R^* ]
	.
	\end{align*}
	This means that the required assumption of Theorem \ref{thm:oracle:bound} is satisfied and we can apply it.

	We restrict the infimumn in \eqref{eq_oracle_in_berns} to $ \rho:= p_0 $ given in \eqref{eq_specific_distribution}.
	Using Lemma \ref{lema_bound_prior_arnak}, we can plug-in the bounds in \eqref{eq_bound_1_1} and \eqref{eq_bound_1_2} into inequality \eqref{eq_oracle_in_berns}, one gets that
	\begin{equation*}
	\mathbb{E} [
	\mathbb{E}_{\theta\sim \hat{\rho}_{\lambda}}
	[ R (\theta) ] ] - R^*
	\leq
	2
	\inf_{\tau \in (0,C_1/2d)} 
	\!
	\left\{  
	C_L	C_{\rm x} 2\varsigma\sqrt{ d }
	+ 
	C_{_{K,C}} \frac{ 4 s^* \log \left(\frac{C_1 }{\varsigma s^*}\right)
		+
		\log(2)
	}{n} \right\}
	,
	\end{equation*}
	and the choice $ \varsigma = ( C_L C_{\rm x} n\sqrt{d})^{-1} $ leads to
	\begin{align*}
	\mathbb{E} [
	\mathbb{E}_{\theta\sim \hat{\rho}_{\lambda}}
	[ R (\theta) ] ] - R^*
	\leq
	2
	\left\{  
	\frac{4}{n} 
	+ 
	C_{_{K,C}} \frac{2 s^* \log \left(\frac{ C_L C_{\rm x} C_1 n\sqrt{d}}{ s^*}\right)
		+	\log(2)
	}{n} \right\}
	\leq
	C_{_{K,C,X,L}}
	\frac{ s^* \log \left(\frac{ n\sqrt{d}}{ s^*}\right)
	}{n}
	,
	\end{align*}
	for some constant $ C_{_{K,C,X,L}} > 0 $ depending only on $ K,C, C_1,C_{\rm x}, C_L $. 
	
	\underline{To obtain the bound in probability:}
	\\
	Fix $\theta\in\Theta$ and apply Bernstein's inequality in Lemma \ref{lemma:bernstein} with $U_i= \ell_i(\theta^*) - \ell_{i}(\theta)  $. We obtain for any $t>0$ that
	$
	\mathbb{E}  {\rm e}^{ t n 
		[R(\theta) - R^* -r(\theta) + r(\theta^*)] }  
	\leq 
	{\rm e}^{ g\left(C t\right) n t^2  {\rm Var} (U_i)}.
	$
	We put $t=\lambda/n$,
	and note that
	\begin{align*}
	{\rm Var} (U_i)
	\leq 
	\mathbb{E} (U_i^2)
	=
	\mathbb{E} \left\{[\ell_i(\theta^*) - \ell_{i}(\theta)]^2 \right\}
	\leq
	C_L	\mathbb{E}	|X_i^\top( \theta - \theta^*) |^2
	\leq 
	C_L	K \left[ R(\theta) - R^* \right]
	\end{align*}
	thanks to Assumption \ref{assum_bernstein}. Thus we get:
	\begin{equation*}
	\mathbb{E}  {\rm e}^{ \lambda 	
		[R(\theta) - R^* -r(\theta) + r(\theta^*)]  } 
	\leq 
	{\rm e}^{ 
		g\left(\frac{\lambda C}{n}\right) \frac{\lambda^2}{n} 
		C_L K \left[ R(\theta) - R^* \right] 
	}
	.
	\end{equation*}
	Reorganize the terms, integrate the modified expression with respect to $ \pi $, and subsequently use Fubini's theorem to derive the following inequality:
	\begin{equation*}
	\mathbb{E}  \mathbb{E}_{\theta\sim\pi} 
	\left( {\rm e}^{\lambda 
		\left\{ \left[ 1- C_L K g\left(\frac{\lambda C}{n}\right) \frac{\lambda}{n} 
		\right]
		\left[ R(\theta) - R^* \right] -r(\theta) + r(\theta^*)\right\} } \right)
	\leq 1.
	\end{equation*}
	and we apply Lemma~\ref{lemma:dv} and multiply both sides by $ \varepsilon >0 $ to get:
	\begin{equation*}
	\mathbb{E} \left[ {\rm e}^{ 
		\lambda 
		\sup_{\rho\in\mathcal{P}(\Theta)} 
		\left\{
		\mathbb{E}_{\theta\sim\rho}
		\left\{
		\left[ 1- C_L K g\left(\frac{\lambda C}{n}\right) \frac{\lambda}{n} \right] [R(\theta) - R^*] -r(\theta) + r(\theta^*) \right\}
		-	
		\mathcal{K}(\rho\|\pi) - \log(1/\varepsilon)
		\right\} 
	} \right] 
	\leq 
	\varepsilon
	.
	\end{equation*} 
	Uses Chernoff bound  to get
	\begin{equation*}
	\mathbb{P} \left[ \sup_{\rho\in\mathcal{P}(\Theta)} \lambda \mathbb{E}_{\theta\sim\rho}
	\left\{
	\left[ 1- C_L K g\left(\frac{\lambda C}{n}\right) \frac{\lambda}{n} \right] 
	[R(\theta) - R^* ] -r(\theta) + r(\theta^*) 
	\right\}
	-	\mathcal{K}(\rho\|\pi)  > \log\frac{1}{\varepsilon} \right] 
	\leq 
	\varepsilon.
	\end{equation*} 
	Take the complement and rearranging terms give, with probability at least $ 1- \varepsilon $
	\begin{equation*}
	\mathbb{E}_{\theta\sim\rho} 
	\left[ 1- C_L K g\left(\frac{\lambda C}{n}\right) \frac{\lambda}{n} \right] 
	[R(\theta) - R^* ] 
	\leq
	\mathbb{E}_{\theta\sim\rho}[ r(\theta) -  r(\theta^*)] +  \frac{	\mathcal{K}(\rho\|\pi) + \log\frac{1}{\varepsilon}}{\lambda} 
	.
	\end{equation*}
	using Lemma \ref{lemma:dv} again, we get that with probability at least $ 1- \varepsilon $
	\begin{multline}
	\label{eq_pacb_fast_1}
	\mathbb{E}_{\theta \sim \hat{\rho}_\lambda}
	\left[ 1- C_L K g\left(\frac{\lambda C}{n}\right) \frac{\lambda}{n} \right] 
	[ R(\theta) - R^* ] 
	\leq
	\inf_{\rho\in\mathcal{P}(\Theta)}  \left\{ 
	\mathbb{E}_{\theta\sim\rho}[ r(\theta) -  r(\theta^*)] + 
	\frac{	\mathcal{K}(\rho\|\pi) + \log\frac{1}{\varepsilon}}{\lambda} 
	\right\} 
	.
	\end{multline}	
	Now, with exactly the same argument but with an application to $ -U_i $ rather than $ U_i $, one obtains that  with probability at least $ 1- \varepsilon $
	\begin{equation}
	\label{eq_pacb_fast_2}
	\mathbb{E}_{\theta\sim\rho}[ r(\theta) -  r(\theta^*)]
	\leq
	\mathbb{E}_{\theta\sim\rho}
	\left[ 1+ C_L K g\left(\frac{\lambda C}{n}\right) \frac{\lambda}{n} \right] 
	[ R(\theta) - R^* ] 
	+ 
	\frac{	\mathcal{K}(\rho\|\pi) + \log\frac{1}{\varepsilon}}{\lambda} 
	.
	\end{equation}
	Combining \eqref{eq_pacb_fast_1} and \eqref{eq_pacb_fast_2}, we get that
	with probability at least $ 1- 2\varepsilon $
	\begin{multline*}
	\mathbb{E}_{\theta \sim \hat{\rho}_\lambda}
	\left[ 1- C_L K g\left(\frac{\lambda C}{n}\right) \frac{\lambda}{n} \right] 
	[ R(\theta) - R^* ] 
	\\
	\leq
	\inf_{\rho\in\mathcal{P}(\Theta)}  \left\{ 
	\mathbb{E}_{\theta\sim\rho}
	\left[ 1+ C_L K g\left(\frac{\lambda C}{n}\right) \frac{\lambda}{n} \right] 
	[ R(\theta) - R^* ] 
	+ 
	2 \frac{	\mathcal{K} (\rho\|\pi) + \log\frac{1}{\varepsilon}}{\lambda} 
	\right\} 
	.
	\end{multline*} 
	From this point, we assume \(\lambda\) is such that \(\left[ 1 - C_L K g\left(\frac{\lambda C}{n}\right) \frac{\lambda}{n} \right] > 0\). As a result,
	\begin{align*}
	\mathbb{E}_{\theta \sim \hat{\rho}_\lambda}
	[ R(\theta) - R^* ] 
	\leq
	\inf_{\rho\in\mathcal{P}(\Theta)}  \left\{ 
	\mathbb{E}_{\theta\sim\rho}
	\frac{ \left[ 1+ C_L K g\left(\frac{\lambda C}{n}\right) \frac{\lambda}{n} \right]  }{ \left[ 1- C_L K g\left(\frac{\lambda C}{n}\right) \frac{\lambda}{n} \right]  }
	[ R(\theta) - R^*] 
	+ 
	2 \frac{ \mathcal{K} (\rho\|\pi) + \log\frac{1}{\varepsilon}}{ \lambda \left[ 1- C_L K g\left(\frac{\lambda C}{n}\right) \frac{\lambda}{n} \right]  } 
	\right\} 
	.
	\end{align*}  
	In particular, take $\lambda = n/\max(2 C_L K,C) $. One has that: $\lambda \leq n/(2C_L K) \Rightarrow K C_L \lambda / n \leq 1/2$ and $\lambda \leq n/C \Rightarrow g(\lambda C/n) \leq g(1) \leq 1 $, so
	$$  
	1 - C_L K g\left(\frac{\lambda C}{n}\right) \frac{\lambda}{n} 
	\geq \frac{1}{2} ; \quad
	1 + C_L K g\left(\frac{\lambda C}{n}\right) \frac{\lambda}{n} 
	\leq \frac{3}{2} 
	.
	$$
	Thus, with probability at least $ 1- 2\varepsilon $, 
	\begin{align}
	\label{eq_pacb_fast_3}
	\mathbb{E}_{\theta \sim \hat{\rho}_\lambda}
	[ R(\theta) - R(\theta^*)] 
	\leq
	\inf_{\rho\in\mathcal{P}(\Theta)}  \left\{ 
	\mathbb{E}_{\theta\sim\rho}
	3
	[ R(\theta) - R(\theta^*)] 
	+ 
	4  \frac{ \mathcal{K} (\rho\|\pi) + \log\frac{1}{\varepsilon} }{ n / \max(2C_L K,C) } 
	\right\} 
	.
	\end{align} 
	We restrict the infimumn in \eqref{eq_pacb_fast_3} to $ \rho:= p_0 $ given in \eqref{eq_specific_distribution}.
	Using Lemma \ref{lema_bound_prior_arnak}, we can plug-in the bounds in \eqref{eq_bound_1_1} and \eqref{eq_bound_1_2} into inequality \eqref{eq_pacb_fast_3}, one gets that
	\begin{equation*}
	\mathbb{E}_{\theta\sim \hat{\rho}_{\lambda}}
	[ R (\theta) ]  - R^*
	\leq
	\inf_{\varsigma \in (0,C_1/2d)} 
	\!
	\left\{  
	6 C_L C_{\rm x} \varsigma\sqrt{ d }
	+ 
	4 C_{_{K,C}} \frac{ 4 s^* \log \left(\frac{C_1 }{\varsigma s^*}\right)
		+
		\log(2) + \log\frac{1}{\varepsilon}
	}{n} \right\}
	,
	\end{equation*}
	and the choice $ \varsigma = (C_L C_{\rm x} n\sqrt{d})^{-1} $ leads to
	\begin{align*}
	\mathbb{E}_{\theta\sim \hat{\rho}_{\lambda}}
	[ R (\theta) ]  - R^*
	\leq
	\frac{6}{n}
	+ 
	4 C_{_{K,C}} \frac{ 4 s^* \log \left(\frac{C_1 C_L C_{\rm x} n \sqrt{d} }{ s^*}\right)
		+
		\log(2) + \log\frac{1}{\varepsilon}
	}{n} 
	\leq
	\mathfrak{C}
	\frac{ s^* \log \left(\frac{ n\sqrt{d}}{ s^*}\right) + \log (1/\varepsilon) 
	}{n}
	,
	\end{align*}
	for some constant $ 	\mathfrak{C}> 0 $ depending only on $ K,C,C_{\rm x}, C_1, C_L $. 	
	The proof is completed.
	
\end{proof}

\begin{proof}[\bf Proof of Proposition \ref{propo_fastrate}]
	
	From Assumption \ref{assum_bernstein}, we have that
	\begin{equation*}
	\mathbb{E} \{ 
	|  X^\top (\theta - \theta^*) |^2 \}
	\leq 
	\mathbb{E} K [ R (\theta) - R(\theta^*) ]
	.
	\end{equation*}
	Integrating with respect ot $ \hat{\rho}_{\lambda} $ and using Fubini's theorem, we gets that 
	\begin{equation*}
	\mathbb{E} 	\mathbb{E}_{\theta\sim \hat{\rho}_{\lambda}} \{ 
	|  X^\top (\theta - \theta^*) |^2 \}
	\leq 
	\mathbb{E} 	\mathbb{E}_{\theta\sim \hat{\rho}_{\lambda}} K [ R (\theta) - R(\theta^*) ]
	,
	\end{equation*}
	using now the results from Theorem \ref{thm_main1}, we obtain that
	\begin{equation*}
	\mathbb{E} 	\mathbb{E}_{\theta\sim \hat{\rho}_{\lambda}} \{ 
	|  X^\top (\theta - \theta^*) |^2 \}
	\leq
	C_{_{K,C,X}}
	\frac{ s^* \log \left(\frac{ n\sqrt{d}}{ s^*}\right) 
	}{n}
	,
	\end{equation*}
	for some constant $ C_{_{K,C,X}} > 0 $ depending only on $ K,C,C_{\rm x} $. 	
	
	With a similar argument but using now Assumption \ref{assum_eigen}, 
	\begin{equation*}
	\mathbb{E} 
	\mathbb{E}_{\theta\sim \hat{\rho}_{\lambda}} \{ 
	|  \theta - \theta^* |^2 \}
	\leq
	C_{_{\kappa,K,C,X}}
	\frac{ s^* \log \left(\frac{ n\sqrt{d}}{ s^*}\right)
	}{n}
	,
	\end{equation*}
	for some constant $ C_{_{\kappa,K,C,X}} > 0 $ depending only on $ \kappa, K,C,C_{\rm x} $. 	
\end{proof}

\subsection{Proof for fast rate with heavy-tailed loss}

\begin{proof}[\bf Proof of Theorem \ref{thm_heavy_tailed_loss}]
	
	Let's define the following
	random variables
	\[
	T_{i} 
	= 
	(Y_i - \exp (X_i^\top \theta^*) )^2
	- (Y_i - \exp (X_i^\top \theta) )^2.
	\]
	Note that these variables are independent. We first check that the
	variables $T_i$
	satisfy the assumptions of Lemma~\ref{lemmemassart}, in order to apply
	this lemma.
	From Assumption \ref{assum_heavy_tailed}, we have that
	\begin{align*}
	\sum_{i=1}^{n} \mathbb{E}[T_{i}^{2}] 
	\leq 
	n H_1	\left[ R(\theta) - R(\theta^*)\right]
	=:
	v
	,
	\end{align*}
	and, for any integer $k\geq3$,
	that
$
	\sum_{i=1}^{n} \mathbb{E}\left[(T_{i})^{k}\right]
	\leq 
	n H_1 [ R (\theta) - R(\theta^*) ] \frac{k! H_2^{k-2}}{2}
	,
$
	with $ w:= H_2 $. Next, for any $\lambda\in(0,n/w)$, applying Lemma \ref{lemmemassart}
	with $\zeta=\lambda/n$ gives
	\[
	\mathbb{E} \exp\left[\lambda
	\Bigl( R(\theta)-R(\theta^*)-r(\theta)+r(\theta^*)\Bigr)\right]
	\leq
	\exp\left[
	\frac{v\lambda^{2}}{2n^{2}(1-\frac{w\lambda}{n})} 
	\right].
	\]
	For any $\varepsilon
	>0$, the last display yields
	\[
	\mathbb{E} \exp\left[
	\Bigl( \lambda - 	\frac{ H_1 \lambda^{2}}{ 2 ( n - H_2 \lambda )}  \Bigr)
	\Bigl( 
	R(\theta) - R(\theta^*) \Bigr)
	+\lambda\Bigl( -r(\theta) + r(\theta^*) \Bigr)
	- \log\frac{2}{\varepsilon} \right] \leq\frac{\varepsilon}{2}.
	\]
	Integrating w.r.t. the probability distribution $ \pi(.) $, and then Fubini's theorem gives
	\begin{align*}
	&	\mathbb{E} \int \exp\Biggl[\alpha
	\Bigl( R(\theta) - R(\theta^*) \Bigr)
	+\lambda\Bigl( -r(\theta) + r(\theta^*) \Bigr)
	- \log\frac{2}{\varepsilon}\Biggr]  \pi (d M)   \hspace*{1.5cm}
	\\
	= & \mathbb{E} \int \exp  \left\lbrace   \alpha    \Bigl( R(\theta) - R(\theta^*) \Bigr)
	+\lambda\Bigl( -r(\theta) + r(\theta^*) \Bigr)               - \log \left[\frac{d\hat{\rho}_{\lambda}}{d \pi} (\theta)  \right]
	- \log\frac{2}{\varepsilon}
	\right\rbrace
	\hat{\rho}_{\lambda}(d M)
	\leq 
	\frac{\varepsilon}{2}
	,
	\end{align*}
	where $ \alpha = \lambda	\Bigl( 1 - 	\frac{ H_1 \lambda}{ 2 ( n - H_2 \lambda )}  \Bigr) $. 
	Jensen's inequality yields
	\[
	\mathbb{E} \exp\Biggl[
	\alpha
	\left( \int R d \hat{\rho}_{\lambda} - R(\theta^*) \right)
	+
	\lambda\left( -\int r d \hat{\rho}_{\lambda} + r(\theta^*) \right)
	- 
	\mathcal{K}(\hat{\rho}_{\lambda}, \pi)
	- 
	\log\frac{2}{\varepsilon}\Biggr] 
	\leq
	\frac{\varepsilon}{2}
	.
	\]
	Now, using the basic inequality $\exp(x) \geq
	\mathbf{1}_{\mathbb{R}_{+}}(x)$, we get
	\[
	\mathbb{P}\Biggl\{ \Bigr[ 
	\alpha
	\Bigl(\int R d\hat{\rho}_{\lambda} - R(\theta^*)\Bigr)
	+
	\lambda\Bigl(-\int r d\hat{\rho}_{\lambda} + r(\theta^*) \Bigr)
	- 
	\mathcal{K}(\hat{\rho}_{\lambda}, \pi)
	- \log\frac{2}{\varepsilon}\Bigr] 
	\geq 0
	\Biggr\} 
	\leq
	\frac{\varepsilon}{2}
	,
	\]
	reorganize the terms to get
	\begin{equation*}
	\mathbb{P}\Biggl\{ \int R d\hat{\rho}_{\lambda} - R(\theta^*)
	\leq
	\frac{ \int r d\hat{\rho}_{\lambda} - r(\theta^*) +
		\frac{1}{\lambda}\left[\mathcal{K}(\hat{\rho}_{\lambda}, \pi)
		+ \log\frac{2}{\varepsilon}\right] } {\frac{\alpha}{\lambda} }
	\Biggr\}
	\geq
	1-\frac{\varepsilon}{2}.
	\end{equation*}
	Using Donsker and Varadhan's variational inequality, Lemma \ref{lemma:dv},
	we get
	%
	\begin{equation}\label{interm3bis}
	\mathbb{P}\Biggl\{ \int R d\hat{\rho}_{\lambda} - R(\theta^*)
	\leq
\inf_{\rho\in \mathcal{P}(\Theta)}
 \frac{ \int r
		d\rho- r(\theta^*) +
		\frac{1}{\lambda}\left[\mathcal{K}(\rho, \pi)
		+ \log\frac{2}{\varepsilon}\right] } {\frac{\alpha}{\lambda} } \Biggr\}
	\geq1-\frac{\varepsilon}{2}
	.
	\end{equation}

	We now want to bound from above $ r(\theta) - r(\theta^*)$ by
	$R(\theta)- R(\theta^*)$. We can use Lemma~\ref{lemmemassart} again, to
	$\tilde{T}_i(\theta) = - T_i(\theta)$ and similar computations yield
	successively
	\[
	\mathbb{E} \exp\left[\lambda\Bigl( R(\theta^*) - R(\theta) +
	r(\theta) - r(\theta^*) \Bigr)\right] \leq
	\exp\left[\frac{v\lambda^{2}}{2n^{2}(1-\frac{w\lambda}{n})}\right],
	\]
	and so for any (data-dependent) $\rho$,
	\[
	\mathbb{E} \exp\Biggl[\beta
	\left(-\int Rd\rho+ R(\theta^*) \right)
	+ \lambda\left( \int r d\rho- r(\theta^*) \right) -
	\mathcal{K}(\rho, \pi) - \log\frac{2}{\varepsilon}\Biggr] \leq
	\frac{\varepsilon}{2},
	\]
	where $ \beta = \lambda	\Bigl( 1 + 	\frac{ H_1 \lambda}{ 2 ( n - H_2 \lambda )}  \Bigr) $.
	So:
	\begin{equation}
	\label{interm4} \mathbb{P}\Biggl\{ \int rd\rho- r(\theta^*)
	\leq\frac{\beta}{\lambda} \left[\int
	Rd\rho- R(\theta^*) \right] + \frac{1}{\lambda}\left[
	\mathcal{K}(\rho, \pi) + \log\frac{2}{\varepsilon} \right]
	\Biggr\}\geq1 - \frac{\varepsilon}{2}.
	\end{equation}
	Combining (\ref{interm4}) and (\ref{interm3bis}) with a union bound
	argument gives the general PAC-Bayesian bound
	%
	\begin{equation}
	\label{PAC_bound_heavytailed}
	\mathbb{P}\Biggl\{\!  \int R d\hat{\rho}_{\lambda} - R(\theta^*)
	\leq
\inf_{\rho\in \mathcal{P}(\Theta) } 
\frac{ \beta[\int Rd\rho-
		R(\theta^*) ] + 2 [
		\mathcal{K}(\rho, \pi) + \log\frac{2}{\varepsilon} ] } {
		\alpha }\! \Biggr\}
	\geq
	1-\varepsilon.
	\end{equation}
	We restrict the infimumn in \eqref{PAC_bound_heavytailed} to $ \rho:= p_0 $ given in \eqref{eq_specific_distribution}.
Using Lemma \ref{lema_bound_prior_arnak}, we can plug-in the bounds in \eqref{eq_bound_1_1} and \eqref{eq_bound_1_2} into inequality \eqref{PAC_bound_heavytailed}, one gets that	
	\begin{equation*}
\mathbb{P}\Biggl\{\!  \int R d\hat{\rho}_{\lambda} - R(\theta^*)
\leq
	\inf_{\varsigma \in (0,C_1/2d)} 
\!
 \frac{ \beta[ C_L	C_{\rm x} 2 \varsigma \sqrt{ d } ] 
 	+ 
 	2 [ 	4 s^* \log \left(\frac{C_1 }{\varsigma s^*}\right)
 	+
 	\log(2) + \log\frac{2}{\varepsilon} ] } {
	\alpha }
\Biggr\}
\geq
1-\varepsilon
.
\end{equation*}
The selection \( \lambda = \frac{n}{H_1 + 2H_2} \) ensures that \( \alpha > 0 \) and \( \lambda < \frac{n}{w} \). With this choice, it follows that \( \alpha/ \beta \leq 3 \); together with the choice $ \varsigma = ( C_L C_{\rm x} n\sqrt{d})^{-1} $ leads to the conclusion that
	\begin{equation*}
\mathbb{P}\Biggl\{\!  \int R d\hat{\rho}_{\lambda} - R(\theta^*)
\leq
\frac{3}{n}
	+ 
	4 (H_1 + 2H_2) \frac{ 4 s^* \log \left(\frac{C_1 C_L	 C_{\rm x} n\sqrt{d} }{ s^*}\right)
	+	\log(2) + \log\frac{2}{\varepsilon}}{ n }
\Biggr\}
\geq
1-\varepsilon
.
\end{equation*}	
The proof is completed.
\end{proof}

\subsection{Lemmas}

First we state a general PAC-Bayesian relative bound in expectation. See Theorem 4.3 in \cite{alquier2024user}.
\begin{theorem}[Theorem 4.3 in \cite{alquier2024user}]
	\label{thm:oracle:bound}
	Assume Assumption \ref{assum_bernstein} is satisfied. Take $\lambda= n/C_{_{K,C}}$, $ C_{_{K,C}} : = \max(2K,C) $, we have:
	\begin{equation}
	\label{eq_oracle_in_berns}
	\mathbb{E} \mathbb{E}_{\theta\sim\hat{\rho}_{\lambda}} [R(\theta) ]- R^*
	\leq 
	2  \inf_{\rho\in\mathcal{P}(\Theta)}  \left\{  \mathbb{E}_{\theta\sim\rho} [R(\theta) ]- R^* 
	+ 
	C_{_{K,C}}	\frac{ \mathcal{K}(\rho\| \pi) }{n} \right\}.
	\end{equation}
\end{theorem}

\begin{lemma}[Hoeffding's inequality]
	\label{lemma:hoeffding}
	Let $U_1, \dots, U_n$ be a set of independent random variables, where each variable takes values from the interval $[a, b]$. Then, we have for any $t>0$ that
	$$ \mathbb{E} 
	{\rm e}^{t \sum_{i=1}^n [ U_i - \mathbb{E}(U_i)]}  \leq {\rm e}^{\frac{n t^2 (b-a)^2}{8}}. $$
\end{lemma}
\begin{proof}
	The proof can be found for example in Chapter 2 of \cite{MR2319879}.
\end{proof}

\begin{lemma}[Bernstein's inequality]
	\label{lemma:bernstein}
	Let the function $g$ define as $g(0) = 1$ and, for $x\neq 0$,
	$ g(x) = \frac{{\rm e}^x - 1 - x}{x^2}. $
	Let $U_1,\dots,U_n$ be i.i.d random variables such that $\mathbb{E}(U_i)$ is well defined and $U_i- \mathbb{E}(U_i) \leq C$ almost surely for some $C\in\mathbb{R}$. Then
	$$
	\mathbb{E}\left( {\rm e}^{t \sum_{i=1}^n [U_i - \mathbb{E}(U_i) ] } \right)
	\leq {\rm e}^{ g\left(C t\right) n t^2 {\rm Var}(U_i) }.
	$$
\end{lemma}
\begin{proof}
	For a proof, see Theorem 5.2.1 in~\cite{catoni2004statistical}.
\end{proof}

\begin{lemma}[Donsker and Varadhan's variational formula, \cite{catonibook}]
	\label{lemma:dv}
	For any measurable, bounded function $h:\Theta\rightarrow\mathbb{R}$ we have:
	\begin{equation*}
	\log \mathbb{E}_{\theta\sim\pi}\left[{\rm e}^{h(\theta)} \right] =\sup_{\rho\in\mathcal{P}(\Theta)}\Bigl[\mathbb{E}_{\theta\sim\rho}[h(\theta)] -	\mathcal{K} (\rho\|\pi)\Bigr].
	\end{equation*}
	Moreover, the supremum with respect to $\rho$ in the right-hand side is
	reached for the Gibbs measure
	$\pi_{h}$ defined by its density with respect to $\pi$
	\begin{equation*}
	\frac{{\rm d}\pi_{h}}{{\rm d}\pi}(\theta) =  \frac{{\rm e}^{h(\theta)}}
	{ \mathbb{E}_{\vartheta\sim\pi}\left[{\rm e}^{h(\vartheta)} \right] }.
	\end{equation*}
\end{lemma}

We define the following distribution as a translation of the prior $ \pi $,
\begin{equation}
\label{eq_specific_distribution}
p_0(\theta) 
\propto 
\pi (\theta - \theta^*)\mathbf{1}_{
	\{ \| \theta - \theta^*\|_1 \leq 2d\varsigma \} 
}
.
\end{equation}

\begin{lemma}
	\label{lema_bound_prior_arnak}
	Let $p_0 $ be the probability measure defined by (\ref{eq_specific_distribution}). If
	$d\geq 2$ then
	$
	\int_\Omega \| \theta- \theta^* \|^2 p_0({\rm d} \theta)
	\leq
	4d\varsigma^2 
	,
	$
	and
	$
	\mathcal{K}(p_0 \| \pi)
	\leq
	4 s^* \log \left(\frac{C_1 }{\varsigma s^*}\right)
	+
	\log(2)
	.
	$
\end{lemma}
\begin{proof}
	The proof can be found in \cite{mai2023high}, which utilizes results from \cite{dalalyan2012mirror}.
\end{proof}

we state a version of Bernstein's inequality  from~\cite{MR2319879}
(Inequality 2.21, page 24).

\begin{lemma}
	\label{lemmemassart} Let $ U_{1}, \ldots,  U_{n} $ be independent real
	valued random variables and assume that there are two constants
	$v$ and $w$ such that
$
	\sum_{i=1}^{n} \mathbb{E}[ U_{i}^{2}] \leq v
	,
$
	and for all integers $k\geq3$,
$
	\sum_{i=1}^{n} \mathbb{E}\left[( U_{i})^{k}_+ \right] \leq v\frac
	{k!w^{k-2}}{2}.
$
	Then, for any $\zeta\in(0,1/w)$,
	\[
	\mathbb{E}
	\exp\left[\zeta\sum_{i=1}^{n}\left[ U_{i} - \mathbb{E}( U_{i})\right]
	\right]
	\leq\exp\left(\frac{v\zeta^{2}}{2(1-w\zeta)} \right) .
	\]
\end{lemma}


\begin{thebibliography}{}
	
	\bibitem[Abramovich and Grinshtein, 2016]{abramovich2016model}
	Abramovich, F. and Grinshtein, V. (2016).
	\newblock Model selection and minimax estimation in generalized linear models.
	\newblock {\em IEEE Transactions on Information Theory}, 62(6):3721--3730.
	
	\bibitem[Abramovich and Grinshtein, 2018]{abramovich2018high}
	Abramovich, F. and Grinshtein, V. (2018).
	\newblock High-dimensional classification by sparse logistic regression.
	\newblock {\em IEEE Transactions on Information Theory}, 65(5):3068--3079.
	
	\bibitem[Abramovich et~al., 2007]{abramovich2007optimality}
	Abramovich, F., Grinshtein, V., and Pensky, M. (2007).
	\newblock On optimality of bayesian testimation in the normal means problem.
	\newblock {\em Annals of Statistics}, 35(5):2261.
	
	\bibitem[Alquier, 2024]{alquier2024user}
	Alquier, P. (2024).
	\newblock {User-friendly introduction to PAC-Bayes bounds}.
	\newblock {\em Foundations and Trends{\textregistered} in Machine Learning},
	17(2):174--303.
	
	\bibitem[Alquier and Biau, 2013]{alquier2013sparse}
	Alquier, P. and Biau, G. (2013).
	\newblock Sparse single-index model.
	\newblock {\em J. Mach. Learn. Res.}, 14:243--280.
	
	\bibitem[Alquier et~al., 2019]{alquier2019estimation}
	Alquier, P., Cottet, V., and Lecu{\'e}, G. (2019).
	\newblock {Estimation bounds and sharp oracle inequalities of regularized
		procedures with Lipschitz loss functions}.
	\newblock {\em The Annals of Statistics}, 47(4):2117 -- 2144.
	
	\bibitem[Alquier and Lounici, 2011]{alquier2011PAC}
	Alquier, P. and Lounici, K. (2011).
	\newblock P{AC}-{B}ayesian bounds for sparse regression estimation with
	exponential weights.
	\newblock {\em Electron. J. Stat.}, 5:127--145.
	
	\bibitem[Alquier et~al., 2016]{alquier2016}
	Alquier, P., Ridgway, J., and Chopin, N. (2016).
	\newblock On the properties of variational approximations of gibbs posteriors.
	\newblock {\em Journal of Machine Learning Research}, 17(236):1--41.
	
	\bibitem[Bissiri et~al., 2016]{bissiri2013general}
	Bissiri, P.~G., Holmes, C.~C., and Walker, S.~G. (2016).
	\newblock A general framework for updating belief distributions.
	\newblock {\em Journal of the Royal Statistical Society Series B: Statistical
		Methodology}, 78(5):1103--1130.
	
	\bibitem[Blaz{\`e}re et~al., 2014]{blazere2014oracle}
	Blaz{\`e}re, M., Loubes, J.-M., and Gamboa, F. (2014).
	\newblock Oracle inequalities for a group lasso procedure applied to
	generalized linear models in high dimension.
	\newblock {\em IEEE Transactions on Information Theory}, 60(4):2303--2318.
	
	\bibitem[Boucheron et~al., 2013]{boucheron2013concentration}
	Boucheron, S., Lugosi, G., and Massart, P. (2013).
	\newblock {\em Concentration inequalities: A nonasymptotic theory of
		independence}.
	\newblock Oxford University Press, Oxford.
	
	\bibitem[B\"uhlmann and van~de Geer, 2011]{buhlmann_vandegeer}
	B\"uhlmann, P. and van~de Geer, S. (2011).
	\newblock {\em Statistics for high-dimensional data: Methods, theory and
		applications}.
	\newblock Springer Series in Statistics. Springer, Heidelberg.
	
	\bibitem[Carvalho et~al., 2010]{carvalho2010horseshoe}
	Carvalho, C.~M., Polson, N.~G., and Scott, J.~G. (2010).
	\newblock The horseshoe estimator for sparse signals.
	\newblock {\em Biometrika}, 97(2):465--480.
	
	\bibitem[Castillo and Mismer, 2018]{castillo2018empirical}
	Castillo, I. and Mismer, R. (2018).
	\newblock Empirical {B}ayes analysis of spike and slab posterior distributions.
	\newblock {\em Electronic Journal of Statistics}, 12:3953--4001.
	
	\bibitem[Castillo et~al., 2015]{castillo2015bayesian}
	Castillo, I., Schmidt-Hieber, J., and {van der Vaart}, A. (2015).
	\newblock Bayesian linear regression with sparse priors.
	\newblock {\em The Annals of Statistics}, 43(5):1986--2018.
	
	\bibitem[Castillo and van~der Vaart, 2012]{castillo2012needles}
	Castillo, I. and van~der Vaart, A. (2012).
	\newblock Needles and straw in a haystack: Posterior concentration for possibly
	sparse sequences.
	\newblock {\em The Annals of Statistics}, pages 2069--2101.
	
	\bibitem[Catoni, 2004]{catoni2004statistical}
	Catoni, O. (2004).
	\newblock {\em Statistical learning theory and stochastic optimization}, volume
	1851 of {\em Saint-Flour Summer School on Probability Theory 2001 (Jean
		Picard ed.), Lecture Notes in Mathematics}.
	\newblock Springer-Verlag, Berlin.
	
	\bibitem[Catoni, 2007]{catonibook}
	Catoni, O. (2007).
	\newblock {\em {PAC}-{B}ayesian supervised classification: the thermodynamics
		of statistical learning}.
	\newblock IMS Lecture Notes---Monograph Series, 56. Institute of Mathematical
	Statistics, Beachwood, OH.
	
	\bibitem[Dalalyan and Tsybakov, 2008]{dalalyan2008aggregation}
	Dalalyan, A. and Tsybakov, A.~B. (2008).
	\newblock Aggregation by exponential weighting, sharp {PAC}-{B}ayesian bounds
	and sparsity.
	\newblock {\em Machine Learning}, 72(1-2):39--61.
	
	\bibitem[Dalalyan, 2017]{dalalyan2017theoretical}
	Dalalyan, A.~S. (2017).
	\newblock Theoretical guarantees for approximate sampling from smooth and
	log-concave densities.
	\newblock {\em Journal of the Royal Statistical Society: Series B (Statistical
		Methodology)}, 3(79):651--676.
	
	\bibitem[Dalalyan and Tsybakov, 2012a]{dalalyan2012mirror}
	Dalalyan, A.~S. and Tsybakov, A. (2012a).
	\newblock Mirror averaging with sparsity priors.
	\newblock {\em Bernoulli}, 18(3):914--944.
	
	\bibitem[Dalalyan and Tsybakov, 2012b]{dalalyan2012sparse}
	Dalalyan, A.~S. and Tsybakov, A.~B. (2012b).
	\newblock Sparse regression learning by aggregation and langevin monte-carlo.
	\newblock {\em Journal of Computer and System Sciences}, 78(5):1423--1443.
	
	\bibitem[Durmus and Moulines, 2019]{durmus2019high}
	Durmus, A. and Moulines, E. (2019).
	\newblock High-dimensional {B}ayesian inference via the unadjusted langevin
	algorithm.
	\newblock {\em Bernoulli}, 25(4A):2854--2882.
	
	\bibitem[D’Angelo and Canale, 2023]{dAngelo2023efficient}
	D’Angelo, L. and Canale, A. (2023).
	\newblock Efficient posterior sampling for bayesian poisson regression.
	\newblock {\em Journal of Computational and Graphical Statistics},
	32(3):917--926.
	
	\bibitem[Elsener and van~de Geer, 2018]{elsener2018robust}
	Elsener, A. and van~de Geer, S. (2018).
	\newblock Robust low-rank matrix estimation.
	\newblock {\em The Annals of Statistics}, 46(6B):3481--3509.
	
	\bibitem[Friedman et~al., 2010]{glmnet}
	Friedman, J., Hastie, T., and Tibshirani, R. (2010).
	\newblock Regularization paths for generalized linear models via coordinate
	descent.
	\newblock {\em Journal of Statistical Software}, 33(1):1--22.
	
	\bibitem[Giraud, 2021]{giraud2021introduction}
	Giraud, C. (2021).
	\newblock {\em Introduction to high-dimensional statistics}.
	\newblock Chapman and Hall/CRC.
	
	\bibitem[Gr{\"u}nwald and Van~Ommen, 2017]{grunwald2017inconsistency}
	Gr{\"u}nwald, P. and Van~Ommen, T. (2017).
	\newblock Inconsistency of {B}ayesian inference for misspecified linear models,
	and a proposal for repairing it.
	\newblock {\em {B}ayesian Analysis}, 12(4):1069--1103.
	
	\bibitem[Guedj, 2019]{guedj2019primer}
	Guedj, B. (2019).
	\newblock A primer on {PAC}-{B}ayesian learning.
	\newblock In {\em S{MF} 2018: {C}ongr\`es de la {S}oci\'{e}t\'{e}
		{M}ath\'{e}matique de {F}rance}, volume~33 of {\em S\'{e}min. Congr.}, pages
	391--413. Soc. Math. France.
	
	\bibitem[Guedj and Alquier, 2013]{guedj2013pac}
	Guedj, B. and Alquier, P. (2013).
	\newblock {PAC-Bayesian} estimation and prediction in sparse additive models.
	\newblock {\em Electron. J. Statist.}, 7:264--291.
	
	\bibitem[Guedj and Robbiano, 2018]{guedj2018pac}
	Guedj, B. and Robbiano, S. (2018).
	\newblock Pac-bayesian high dimensional bipartite ranking.
	\newblock {\em Journal of Statistical Planning and Inference}, 196:70--86.
	
	\bibitem[Hastie et~al., 2009]{hastie2009elements}
	Hastie, T., Tibshirani, R., Friedman, J.~H., and Friedman, J.~H. (2009).
	\newblock {\em The elements of statistical learning: data mining, inference,
		and prediction}, volume~2.
	\newblock Springer.
	
	\bibitem[Jeong and Ghosal, 2021]{jeong2021posterior}
	Jeong, S. and Ghosal, S. (2021).
	\newblock Posterior contraction in sparse generalized linear models.
	\newblock {\em Biometrika}, 108(2):367--379.
	
	\bibitem[Jiang, 2007]{jiang2007bayesian}
	Jiang, W. (2007).
	\newblock Bayesian variable selection for high dimensional generalized linear
	models: Convergence rates of the fitted densities.
	\newblock {\em The Annals of Statistics}, 35(4):1487--1511.
	
	\bibitem[Johnstone and Silverman, 2004]{johnstone2004needles}
	Johnstone, I.~M. and Silverman, B.~W. (2004).
	\newblock Needles and straw in haystacks: {{Empirical Bayes}} estimates of
	possibly sparse sequences.
	\newblock {\em Annals of Statistics}, 32:1594--1649.
	
	\bibitem[Kharabati and Amini, 2023]{kharabati2023variat}
	Kharabati, M. and Amini, M. (2023).
	\newblock Variational inference for sparse poisson regression.
	\newblock {\em arXiv}, 2311.01147.
	
	\bibitem[Knoblauch et~al., 2022]{kno2019}
	Knoblauch, J., Jewson, J., and Damoulas, T. (2022).
	\newblock An optimization-centric view on bayes’ rule: Reviewing and
	generalizing variational inference.
	\newblock {\em Journal of Machine Learning Research}, 23(132):1--109.
	
	\bibitem[Lehman and Archer, 2019]{lehman2019penalized}
	Lehman, R.~R. and Archer, K.~J. (2019).
	\newblock Penalized negative binomial models for modeling an overdispersed
	count outcome with a high-dimensional predictor space: Application predicting
	micronuclei frequency.
	\newblock {\em PloS one}, 14(1):e0209923.
	
	\bibitem[Li et~al., 2022]{li2022heterogeneous}
	Li, S., Wei, H., and Lei, X. (2022).
	\newblock Heterogeneous overdispersed count data regressions via
	double-penalized estimations.
	\newblock {\em Mathematics}, 10(10):1700.
	
	\bibitem[Luu et~al., 2019]{luu2019pac}
	Luu, T.~D., Fadili, J., and Chesneau, C. (2019).
	\newblock {PAC-Bayesian risk bounds for group-analysis sparse regression by
		exponential weighting}.
	\newblock {\em Journal of Multivariate Analysis}, 171:209--233.
	
	\bibitem[Mai, 2024a]{mai2023high}
	Mai, T.~T. (2024a).
	\newblock High-dimensional sparse classification using exponential weighting
	with empirical hinge loss.
	\newblock {\em Statistica Neerlandica}, (accepted):arXiv:2312.12952.
	
	\bibitem[Mai, 2024b]{mai2024sparse}
	Mai, T.~T. (2024b).
	\newblock A sparse pac-bayesian approach for high-dimensional quantile
	prediction.
	\newblock {\em arXiv preprint arXiv:2409.01687}.
	
	\bibitem[Mai and Alquier, 2015]{mai2015}
	Mai, T.~T. and Alquier, P. (2015).
	\newblock A {B}ayesian approach for noisy matrix completion: Optimal rate under
	general sampling distribution.
	\newblock {\em Electron. J. Statist.}, 9(1):823--841.
	
	\bibitem[Massart, 2007]{MR2319879}
	Massart, P. (2007).
	\newblock {\em Concentration inequalities and model selection}, volume 1896 of
	{\em Lecture Notes in Mathematics}.
	\newblock Springer, Berlin.
	\newblock Lectures from the 33rd Summer School on Probability Theory held in
	Saint-Flour, July 6--23, 2003, Edited by Jean Picard.
	
	\bibitem[McAllester, 1998]{McA}
	McAllester, D. (1998).
	\newblock Some {{PAC}}-{B}ayesian theorems.
	\newblock In {\em Proceedings of the Eleventh Annual Conference on
		Computational Learning Theory}, pages 230--234, New York. ACM.
	
	\bibitem[McCullagh and Nelder, 1989]{mccullagh1989generalized}
	McCullagh, P. and Nelder, J.~A. (1989).
	\newblock {\em Generalized linear models}.
	\newblock Monographs on Statistics and Applied Probability. Chapman \& Hall,
	London, 2nd edition.
	
	\bibitem[Mendelson, 2008]{mendelson2008obtaining}
	Mendelson, S. (2008).
	\newblock Obtaining fast error rates in nonconvex situations.
	\newblock {\em Journal of Complexity}, 24(3):380--397.
	
	\bibitem[Peng, 2024]{peng2024oracle}
	Peng, L. (2024).
	\newblock Oracle inequalities for weighted group lasso in high-dimensional
	poisson regression model.
	\newblock {\em Communications in Statistics-Theory and Methods},
	53(19):6891--6917.
	
	\bibitem[Ray and Szab{\'o}, 2022]{ray2022variational}
	Ray, K. and Szab{\'o}, B. (2022).
	\newblock Variational bayes for high-dimensional linear regression with sparse
	priors.
	\newblock {\em Journal of the American Statistical Association},
	117(539):1270--1281.
	
	\bibitem[Richards, 2008]{richards2008dealing}
	Richards, S.~A. (2008).
	\newblock Dealing with overdispersed count data in applied ecology.
	\newblock {\em Journal of Applied Ecology}, 45(1):218--227.
	
	\bibitem[Ridgway et~al., 2014]{ridgway2014pac}
	Ridgway, J., Alquier, P., Chopin, N., and Liang, F. (2014).
	\newblock {PAC-Bayesian AUC classification and scoring}.
	\newblock In Ghahramani, Z., Welling, M., Cortes, C., Lawrence, N., and
	Weinberger, K., editors, {\em Advances in Neural Information Processing
		Systems}, volume~27, pages 658--666. Curran Associates, Inc.
	
	\bibitem[Rivoirard, 2006]{rivoirard2006nonlinear}
	Rivoirard, V. (2006).
	\newblock Nonlinear estimation over weak besov spaces and minimax bayes method.
	\newblock {\em Bernoulli}, 12(4):609--632.
	
	\bibitem[Seeger, 2008]{seeger2008bayesian}
	Seeger, M.~W. (2008).
	\newblock Bayesian inference and optimal design for the sparse linear model.
	\newblock {\em Journal of Machine Learning Research}, 9:759--813.
	
	\bibitem[Shawe-Taylor and Williamson, 1997]{STW}
	Shawe-Taylor, J. and Williamson, R. (1997).
	\newblock A {{PAC}} analysis of a {B}ayes estimator.
	\newblock In {\em Proceedings of the Tenth Annual Conference on Computational
		Learning Theory}, pages 2--9, New York. ACM.
	
	\bibitem[Tutz, 2011]{tutz2011regression}
	Tutz, G. (2011).
	\newblock {\em Regression for categorical data}, volume~34.
	\newblock Cambridge University Press.
	
	\bibitem[Wainwright, 2019]{wainwright2019high}
	Wainwright, M.~J. (2019).
	\newblock {\em High-dimensional statistics: A non-asymptotic viewpoint},
	volume~48.
	\newblock Cambridge university press.
	
	\bibitem[Wang et~al., 2016]{wang2016penalized}
	Wang, Z., Ma, S., Zappitelli, M., Parikh, C., Wang, C.-Y., and Devarajan, P.
	(2016).
	\newblock Penalized count data regression with application to hospital stay
	after pediatric cardiac surgery.
	\newblock {\em Statistical methods in medical research}, 25(6):2685--2703.
	
	\bibitem[Zhang and Jia, 2022]{zhang2022elastic}
	Zhang, H. and Jia, J. (2022).
	\newblock Elastic-net regularized high-dimensional negative binomial
	regression.
	\newblock {\em Statistica Sinica}, 32(1):181--207.
	
	\bibitem[Zhang, 2004]{zhang2004statistical}
	Zhang, T. (2004).
	\newblock Statistical behavior and consistency of classification methods based
	on convex risk minimization.
	\newblock {\em The Annals of Statistics}, 32(1):56--85.
	
	\bibitem[Zilberman and Abramovich, 2024]{zilberman2024high}
	Zilberman, O. and Abramovich, F. (2024).
	\newblock High-dimensional regression with a count response.
	\newblock {\em arXiv preprint arXiv:2409.08821}.
	
\end{thebibliography}
\end{document}